\documentclass[preprint,aps,showpacs]{revtex4}

\usepackage{amsmath}
\usepackage{graphicx}

\newcommand{\ds}{\displaystyle}
\newcommand{\mbf}[1]{\ensuremath{\mathbf{#1}}}
\newcommand{\bds}[1]{\ensuremath{\boldsymbol{#1}}}
\newcommand{\adjoint}[1]{\ensuremath{#1^\dagger}}
\renewcommand{\tensor}[1]{\ensuremath{\overset{=}{\mbf{#1}}}}

\newcommand{\uvec}[1]{\ensuremath{\mbf{\hat{e}}_{#1}}}

\begin{document}
\title{Theory of Synergy between Electron Cyclotron\\ and Lower Hybrid Waves}
\author{R.~J. Dumont}
\email{remi.dumont@cea.fr}
\author{G. Giruzzi}
\affiliation{Association EURATOM-CEA sur la Fusion, CEA/DSM/DRFC\\CEA-Cadarache, 13108 Saint-Paul-lez-Durance (France)}
\date{\today}

\begin{abstract}
A theoretical study of the improvement of the Electron Cyclotron Current Drive (ECCD) efficiency in regimes where most of the current is driven by Lower Hybrid (LH) waves is presented. A perturbation technique is employed to solve the adjoint equation and derive the response function including both collisional and LH effects in the limit where the former dominate. An alternative treatment of the problem, involving a numerical solution of the Langevin equations is proposed to gain insight into the current drive mechanism and confirm the obtained results. The existence of a cross-effect between the two waves is demonstrated and the conditions for the synergy, i.e. significant enhancement of the ECCD efficiency in the presence of LH power, are identified.
\end{abstract}

\pacs{52.35.Hr, 52.35.Mw, 52.55.Wq}

\maketitle
\section{Introduction}
Among the desirable features of a future tokamak-based fusion reactor is its steady-state operation, which implies that the toroidal current has to be totally sustained by non inductive sources\cite{Fisch1987}. Moreover, for ongoing as well as for future experiments, a sharp tailoring of the current profiles is known to have a favorable effect on the plasma confinement. These prerequisites are key elements of the advanced tokamak concept, which relies on the fact that a large fraction of the plasma current is supplied by the bootstrap current, generally triggered and supplemented by injecting neutral beams of radio-frequency (RF) waves in the plasma\cite{Taylor1997}.

A wide class of waves can be launched in the plasma yet, for current drive purposes, the excitation of superthermal electrons has been the most successful method, in accordance with theoretical predictions\cite{Fisch1987}. Lower Hybrid Current Drive (LHCD) is a well-tested and efficient method\cite{Litaudon1996,Barbato1998,Peysson2000}, based on Landau damping of the wave power. Its main drawback is that in the so-called multi-pass regime, the current profile remains difficult to control and calculations still lack full reliability. Note, however, that in very hot plasmas, the wave is expected to be absorbed without relying on any subtle upshift mechanism (single-pass absorption), which makes the LH wave a serious candidate to drive off-axis current in future experiments\cite{Barbato1998}. On the other hand, Electron Cyclotron (EC) waves exploit the resonance with the electron gyro-motion and are characterized by a narrow deposition, thus providing a means to induce a local modification of the current profile. The efficiency, however, is known to be significantly lower than for LH waves\cite{Erckmann1994}.

Owing to these complementary features, combined schemes, in which LH and EC waves are used together, constitute an appealing solution for advanced tokamak discharges. In particular, both waves are expected to exert a cross-effect on superthermal electrons, often and sometimes improperly referred to as the LH-EC synergy effect. The consequences of this cross-effect encompass the possibility to modify the LH superthermal tail, a better absorption of EC waves and an improvement of the ECCD efficiency which may help in controlling the current profile\cite{Fidone1984,Fidone1987,Farina1988,Farina1989,Giruzzi1989}. Self-consistent, dynamical calculations including waves kinetic and heat transport effects have stressed the advantages of such scenarios but owing to the non-linear nature of these simulations, the various effects are difficult to separate\cite{Dumont2000}. Experimentally, a cross-effect has been observed  under certain conditions but the interpretation of these measurements are difficult and have led to contradictory conclusions, due either to their transient character or to very large fast particle losses\cite{Ando1986,Yamamoto1987,Maekawa1993,Colborn1998}. More recently, a hard X-ray camera\cite{Peysson1999} has been employed to analyze the emission of fast electrons in the presence of LH and EC waves during the current flat-top phase on the Tore Supra\cite{Giruzzi2000} and FTU\cite{Pericoli2001} tokamaks. In both experiments, a response of the high energy channels was observed, compatible with the existence of a cross-effect, but due to the relatively low applied EC power, no conclusive observation could be made in terms of ECCD efficiency.

Despite these numerical and experimental results, and owing to the lack of a simple mechanism to explain the cross-effect of both waves on fast electrons, this subject is still being debated. The goal of this paper is thus to study the combined current drive process from a theoretical standpoint and identify the conditions for the existence of a LH-EC synergy. To this aim, the adjoint method, originally proposed by Antonsen and Chu\cite{Antonsen1982} and generalized to the RF current drive problem by Fisch\cite{Fisch1987}, is extended to a situation in which two waves are simultaneously present in the plasma. Another possible method is to solve the Langevin equations, which track individual electron relaxation paths and deduce quantities of interest for the current drive problem by average over statistical realizations\cite{Risken1989}. These two complementary methods are employed to derive the response function, which leads to an estimate of the current drive efficiency, including the cross-effect and therefore the LH-EC synergy.

This article is organized as follows. After the presentation of the kinetic aspect of the problem in section \ref{sec:fast-kinetic}, the adjoint method is employed to compute an approximate expression of the response function for a LHCD plasma in section \ref{sec:adjoint-method}. Section \ref{sec:langevin-equations} is devoted to the study of the dynamics underlying the RF current drive process, through the derivation of the associated Langevin equations. The properties of the response function obtained from these two methods are examined in section \ref{sec:response-function-lhcd}. The practical consequences of the presence of LH waves on the EC Current Drive are discussed in section \ref{sec:lhec-efficiency}, where the synergy effect is demonstrated. Conclusions are drawn is section \ref{sec:conclusions}.


\section{Kinetic modeling of LH+EC current drive}
\label{sec:fast-kinetic}
In the absence of a static parallel electric field and including the effects of Coulomb collisions, LH and EC waves, the kinetic equation can be written as
\begin{equation}
\frac{\partial f}{\partial\tau}-\hat{C}f=\hat{D}_{lh}f+\hat{D}_{ec}f\label{eq:kinetic-general}
\end{equation}

In this expression, $\tau\equiv \nu_e t$ is the time in terms of the collision period $\nu_e^{-1}$, $f$ is the electron distribution function. $\mbf{u}\equiv\mbf{p}/\sqrt{m_eT_e}$ is the normalized momentum.

$\hat{C}$ is the linearized collision operator. In this paper, we shall assume that for any perturbed distribution function written as $f\equiv f_m(1+\alpha)$, the high velocity collision operator $\hat{C}f\equiv\hat{C}(f_m\alpha)$ is linearized according to $\hat{C}(f_m\alpha)=\hat{C}(f_m,f_m\alpha)+\hat{C}(f_m\alpha,f_m)+\hat{C}(f_m\alpha,f_i)$, $f_i$ being the ion distribution function. If $\mu\equiv u_\parallel/u$ refers to the cosine of the pitch angle, it can be expressed as\cite{Fisch1987}

\begin{equation}
\hat{C}f\equiv \frac{2}{u^2}\frac{\partial}{\partial u}\bigg(\frac{1}{u}\frac{\partial f}{\partial u}+f\bigg)+\frac{Z_i+1}{u^3}\frac{\partial}{\partial\mu}(1-\mu^2)\frac{\partial f}{\partial\mu}\label{eq:ccoll-umu}
\end{equation}

where $Z_i$ is the plasma ion charge. 

$\hat{D}_{lh}$ (resp. $\hat{D}_{ec}$) is the quasilinear operator associated to LH (resp. EC) waves, which is related to the corresponding quasilinear diffusion tensor $\tensor{D}_{lh}$ (resp. $\tensor{D}_{ec}$) and flux $\mbf{S}_{lh}$ (resp. $\mbf{S}_{ec}$) by
\begin{equation}
\hat{D}_{lh}f=\frac{\partial}{\partial\mbf{u}}\cdot\tensor{D}_{lh}\cdot\frac{\partial f}{\partial\mbf{u}}=-\frac{\partial}{\partial\mbf{u}}\cdot\mbf{S}_{lh}
\end{equation}
and
\begin{equation}
\hat{D}_{ec}f=\frac{\partial}{\partial\mbf{u}}\cdot\tensor{D}_{ec}\cdot\frac{\partial f}{\partial\mbf{u}}=-\frac{\partial}{\partial\mbf{u}}\cdot\mbf{S}_{ec}
\label{eq:slh}
\end{equation}

The LH wave is absorbed in the plasma by Landau damping, which implies that $\tensor{D}_{lh}$, the corresponding quasilinear diffusion tensor is dominated by its parallel-parallel component\cite{Brambilla1998} and the corresponding quasilinear operator can thus be written under the form
\begin{equation}
\hat{D}_{lh}\equiv \frac{D_{lh}}{\nu_e m_e T_e}\cdot\frac{\partial}{\partial u_{\parallel}}d_{lh}(u_\parallel)\frac{\partial}{\partial u_{\parallel}}\label{eq:dlhql}
\end{equation}

$T_e$ is the local electron temperature, $D_{lh}$ is a constant determined by the wave power, so that $D_{lh,0}\equiv D_{lh}/\nu_e m_e T_e$ quantifies its relative intensity compared to collisions. $d_{lh}\equiv d_{lh}(u_\parallel)$ represents the shape of the diffusion coefficient. Here, we consider a regime in which the electrons undergo the effects of the wave in a region of velocity space bounded by two limits, $u_{\parallel,1}$ and $u_{\parallel,2}$, determined by the propagation properties of the wave\cite{Dumont2000}. The following shape is assumed
\begin{equation}
d_{lh}(u_\parallel)\equiv\left\{
\begin{array}{ll}
A_l\exp\big(-(u_\parallel-u_{\parallel,l})^2/\Delta u_{\parallel,1}^2\big) & \textrm{, $u_{\parallel}<u_{\parallel,1}$}\\
u_{\parallel,1}/u_\parallel & \textrm{, $u_{\parallel,1}\le u_{\parallel}\le u_{\parallel,2}$}\\
A_r\exp\big(-(u_\parallel-u_{\parallel,r})^2/\Delta u_{\parallel,2}^2\big) & \textrm{, $u_{\parallel}>u_{\parallel,2}$}
\end{array}\right.
\label{eq:dlh-qldiff}
\end{equation}

$A_r$, $A_l$, $u_{\parallel,r}$ and $u_{\parallel,l}$ are constants whose values are determined by the requirement that both $d_{lh}$ and $\partial d_{lh}/\partial u_{\parallel}$ be continuous at $u_\parallel=u_{\parallel,1}$ and $u_\parallel=u_{\parallel,2}$.

For the problem under discussion here, a useful form for the distribution function is $f\equiv f_m(1+\phi+\delta\phi)$, where $f_m$ is the Maxwellian. $f_m(1+\phi)$ is the distribution function modified by application of the LH power, solution of
\begin{equation}
\frac{\partial f_m\phi}{\partial\tau}-\hat{C}(f_m\phi)=\hat{D}_{lh}f_m(1+\phi)\label{eq:kinetic-lh}
\end{equation}

Upon subtracting (\ref{eq:kinetic-lh}) from (\ref{eq:kinetic-general}), we obtain the equation for $f_m\delta\phi$, which writes
\begin{equation}
\ds{\frac{\partial f_m\delta\phi}{\partial\tau}-\hat{C}(f_m\delta\phi)=\hat{D}_{ec}f_m(1+\phi+\delta\phi)}+\hat{D}_{lh}(f_m\delta\phi)\label{eq:kinetic-ec}
\end{equation}

or equivalently
\begin{equation}
\frac{\partial f_m\delta\phi}{\partial\tau}-\hat{C}(f_m\delta\phi)-\hat{D}_{lh}(f_m\delta\phi)=-\frac{\partial}{\partial\mbf{u}}\cdot\mbf{S}_{ec}\label{eq:kinetic-ec-relax}
\end{equation}

In this expression, the right term describes the electron excitation caused by the electron cyclotron waves whereas $(\hat{C}+\hat{D}_{lh})(f_m\delta\phi)$ is representative of the relaxation under the combined effect of collisions and LH power.

Recalling that $f_m$ is even for $u_\parallel$, the normalized current associated to $f$ can be written as
\begin{equation}
j=\int d\mbf{u}f_m(\phi+\delta\phi)\equiv j_0+j_1\label{eq:normcurrent}
\end{equation}

where $j_0$ and $j_1$ are
\begin{equation}
j_0\equiv\int d\mbf{u}u_\parallel f_m\phi\textrm{,}\hspace{1cm}j_1\equiv\int d\mbf{u}u_\parallel f_m\delta\phi\label{eq:j0j1}
\end{equation}

Note that up to this point, no approximation other than the fact that $T_e$ and $n_e$, the electron temperature and density, are non-varying on the time-scale of the studied problem has been introduced. To evaluate the driven current, it is usual to resort to a Fokker-Planck code and compute the distribution function solution of Eq.~\ref{eq:kinetic-general}. In this work, however, an analytical approach of the problem has been preferred. Even though it implies certain assumptions, the linearization it is based on has the advantage of allowing a separate treatment of the two waves, which is impracticable in purely numerical calculations. This separation is the key to demonstrating unambiguously the possibility of a synergy effect.

To compute $j_0$ (Eq.~\ref{eq:j0j1}), it is necessary to solve Eq.~\ref{eq:kinetic-lh}. Following Eq.~\ref{eq:slh}, the expression for the LH-induced quasilinear flux is 
\begin{equation}
\mbf{S}_{lh}=-\tensor{D}_{lh}\cdot\frac{\partial f}{\partial\mbf{u}}=-\tensor{D}_{lh}\cdot\frac{\partial f_m(1+\phi+\delta\phi)}{\partial\mbf{u}}
\end{equation}

If $\mbf{S}_{lh}$ is not significantly modified by the presence of EC waves, or equivalently assuming that the shape of the distribution function is mainly determined by the effects of collisions and LH wave, i.e. $|\delta\phi|\ll|\phi|$, Eq.~\ref{eq:kinetic-lh} can be rewritten as
\begin{equation}
\frac{\partial f_m\phi}{\partial\tau}-\hat{C}(f_m\phi)=-\frac{\partial}{\partial\mbf{u}}\cdot\mbf{S}_{lh}
\label{eq:kinetic-lh-approx}
\end{equation}

Introducing $g_0(\mbf{u},\mbf{u}',\tau-\tau')$ the Green function associated with Eq.~\ref{eq:kinetic-lh-approx}, solution of
\begin{equation}
\frac{\partial g_0}{\partial\tau}-\hat{C}g_0=\delta(\mbf{u}-\mbf{u}')\delta(\tau-\tau')
\label{eq:green-g0}
\end{equation}

and defining the steady state response function as
\begin{equation}
\chi_0(\mbf{u})\equiv\int_0^\infty d\tau'\int d\mbf{u'}u'_\parallel g_0(\mbf{u},\mbf{u'},\tau')
\end{equation}

$j_0$ is given by
\begin{equation}
j_0=\int d\mbf{u}\,\mbf{S}_{lh}\cdot\frac{\partial\chi_0}{\partial\mbf{u}}
\label{eq:def-j0}
\end{equation}

Likewise, to compute $j_1$, a similar treatment is applied to Eq.~\ref{eq:kinetic-ec}, introducing $g_1(\mbf{u},\mbf{u}',\tau-\tau')$ the solution of the Green problem for Eq.~\ref{eq:kinetic-ec-relax}. It solves
\begin{equation}
\frac{\partial g_1}{\partial\tau}-\hat{C}g_1-\hat{D}_{lh}g_1=\delta(\mbf{u}-\mbf{u}')\delta(\tau-\tau')
\label{eq:green-g1}
\end{equation}

The associated steady-state response function writes
\begin{equation}
\chi_1(\mbf{u})\equiv\int_0^\infty d\tau'\int d\mbf{u'}u'_\parallel g_1(\mbf{u},\mbf{u'},\tau')
\label{eq:chi1}
\end{equation}
 
and allows to evaluate $j_1$ by using
\begin{equation}
j_1=\int d\mbf{u}\,\mbf{S}_{ec}\cdot\frac{\partial\chi_1}{\partial\mbf{u}}
\label{eq:def-j1}
\end{equation}

This method allows to envision the current drive mechanism as a two steps process:
\begin{enumerate}
\item A drive, whose features are contained in the expressions for the quasilinear fluxes $\mbf{S}_{lh}$ and $\mbf{S}_{ec}$. 
\item A relaxation, which is described by the response functions $\chi_0$ and $\chi_1$. 
\end{enumerate}

The quasilinear fluxes contain the information on the distribution function and their evaluation is a delicate task, generally involving a kinetic code. However, if one is merely interested in an estimation of the current drive efficiency, as long as the interaction is well localized in velocity space, the information on the direction of these fluxes is known to be sufficient\cite{Fisch1987}. Note that this relaxes the assumption employed to derive Eq.~\ref{eq:kinetic-lh-approx} since as a result, only the direction of $\mbf{S}_{lh}$ has to be unchanged by EC waves for the LH efficiency calculation to remain valid. 

The problem now reduces to evaluating the response function and in order to achieve this, two methods are available: the adjoint formalism and the Langevin equations. Both will be discussed for the case of two waves in the next sections.

\section{Adjoint method in the presence of two waves}
\label{sec:adjoint-method}
In this section, the adjoint formalism is extended to the case of two waves, when no static electric field is present, which is relevant for a number of experiments. For the sake of concision, only the significant steps of this method, extensively discussed in Ref.~\onlinecite{Fisch1987} and references therein, will be recalled.

By making use of Eq.~\ref{eq:kinetic-lh-approx} when steady state is attained, the current $j_0$ defined in Eq.~\ref{eq:def-j0} can be written as
\begin{equation}
j_0=-\int d\mbf{u}\,\chi_0 \hat{C}(f_m\phi)\label{eq:j0-inverse}
\end{equation}

Introducing the commutative operation for two functions $\varphi(\mbf{u})$ and $\psi(\mbf{u})$
\begin{equation}
[\varphi,\psi]\equiv\int d\mbf{u}\varphi(\mbf{u})\psi(\mbf{u})\label{eq:eqcomadjoint}
\end{equation}

and defining the adjoint $\adjoint{\hat{D}}$ of an operator $\hat{D}$ as
\begin{equation}
[\varphi,\adjoint{\hat{D}}\psi]=[\hat{D}\varphi,\psi]\label{eq:defadjoint}
\end{equation}

Eq.~\ref{eq:j0-inverse} can be rewritten as
\begin{equation}
j_0=-\int d\mbf{u}f_m\phi\adjoint{\hat{C}}\chi_0
\label{eq:j0-adjoint}
\end{equation}

The adjoint equation is obtained by comparing Eqs.~\ref{eq:j0-adjoint} and \ref{eq:j0j1} and making use of the property $f_m\adjoint{\hat{C}}\psi=\hat{C}(f_m\psi)$. It writes
\begin{equation}
\hat{C}\chi_0=-u_\parallel
\label{eq:adjoint-chi0}
\end{equation}

This equation simply describes the response of a collisional plasma and associated with Eq.~\ref{eq:ccoll-umu}, it leads to the well-known Fisch-Boozer response function 
\begin{equation}
\chi_0=\frac{1}{2(5+Z_i)}u^4\mu
\label{eq:chi-fischboozer}
\end{equation}

In order to compute $j_1$, a similar treatment is applied to Eq.~\ref{eq:kinetic-ec-relax}, introducing the associated response function $\chi_1(\mbf{u})$. By noting that the LH quasilinear diffusion coefficient is self-adjoint, i.e. $\adjoint{\hat{D}}_{lh}=\hat{D}_{lh}$, the adjoint equation takes the form
\begin{equation}
[\hat{C}+\hat{D}_{lh}]\chi_1=-u_\parallel
\label{eq:adjoint-chi1}
\end{equation}

This equation describes $\chi_1$, response function of a plasma in which the LH wave modifies the distribution function, modifying in turn the electron relaxation properties. Physically, it means that these electrons describe a collisional curve in velocity space, carrying elemental current $u_\parallel+\adjoint{\hat{D}}_{lh}\chi_1$ instead of $u_\parallel$. A rigorously equivalent interpretation is that the particles carry $u_\parallel$ as elemental current but describe relaxation curves influenced by the wave.

Under this form, the adjoint equation (\ref{eq:adjoint-chi1}) does not appear to have an analytical solution. A further assumption is to consider that collisions dominate the relaxation process. In other words, in spite of the modification of the dynamics in the parallel direction caused by the LH wave, the relaxation curves remain mostly collisional. This approximation allows one to linearize $\chi_1$, letting $\chi_1\equiv\bar{\chi}+\delta{\chi}$ with $|\delta\chi|\ll|\bar{\chi}|$, the small parameter being  $D_{lh,0}$.

The zero-th order expansion of (\ref{eq:adjoint-chi1}) writes
\begin{equation}
\hat{C}\bar{\chi}=-u_\parallel
\end{equation}

which demonstrates that $\bar{\chi}$ is exactly the Fisch-Boozer response function $\chi_0$.

To first order, we obtain
\begin{equation}
\hat{C}\delta\chi=-\hat{D}_{lh}\chi_0
\end{equation}

which, upon expanding the quasilinear operators and letting $\hat{Z}=(Z_i+1)/2$, yields
\begin{equation}
u\frac{\partial\delta\chi}{\partial u}-\hat{Z}\frac{\partial}{\partial\mu}(1-\mu^2)\frac{\partial\delta\chi}{\partial\mu}=\frac{D_{lh}}{2\nu_e m_e T_e}u^3\frac{\partial}{\partial u_\parallel}d_{lh}(u,\mu)\frac{\partial\chi_0}{\partial u_\parallel}\label{eq:shmaster}
\end{equation}

The associated Green equation writes
\begin{equation}
u\frac{\partial G_\chi}{\partial u}-\hat{Z}\frac{\partial}{\partial\mu}(1-\mu^2)\frac{\partial G_\chi}{\partial\mu}=\frac{\delta(u-u')\delta(\mu-\mu')}{{u'}^2}\label{eq:greengchi}
\end{equation}

Here, $G_\chi(\mbf{u},\mbf{u'})$ is the steady-state Green function of the problem.

The Pitch-angle scattering term in Eq.~\ref{eq:greengchi} suggests the expansion\cite{Rax1989}
\begin{equation}
\delta(\mu'-\mu)=\sum_{l=0}^\infty\frac{(2l+1)}{2}P_l(\mu)P_l(\mu')
\end{equation}

where $(P_l)$ are the Legendre polynomials and applying a variable separation, one obtains
\begin{equation}
G_\chi(\mbf{u},\mbf{u}')=\frac{Y(u-u')}{u^3}\sum_{l=0}^\infty\frac{(2l+1)}{2}\bigg(\frac{u'}{u}\bigg)^{\hat{Z}l(l+1)}P_l(\mu)P_l(\mu')
\end{equation}

where $Y$ is the Heaviside function.

This leads to the solution for $\delta\chi$
\begin{equation}
\delta\chi(u,\mu)=\frac{1}{4(5+Z_i)}\bigg(\frac{D_{lh}}{\nu_e m_e T_e}\bigg)u^4\sum_{l=0}^\infty\frac{(2l+1)}{2}Q_l(u)P_l(\mu)\label{eq:deltachi}
\end{equation}

with
\begin{equation}
Q_l(u)\equiv\int_0^{u}du'\bigg(\frac{u'}{u}\bigg)^{\hat{Z}l(l+1)+4}J_l(u')
\end{equation}

and
\begin{equation}
J_l(u')\equiv \int_{-1}^1d\mu' P_l(\mu')\bigg[3d_{lh}(u',\mu')\mu'(3+{\mu'}^2)+\frac{\partial d_{lh}}{\partial u'_\parallel}u'(3{\mu'}^2+1)\bigg]
\end{equation}

\section{Langevin equations}
\label{sec:langevin-equations}
To track the electrons trajectories on their relaxation paths, a natural and convenient method consists in solving the Langevin equations\cite{Fisch1987,Cadjan1999,Castejon2003}. Besides providing a clear insight in the dynamics underlying the relaxation process\cite{GarciaPalacios1998}, they can be used to compute the response function $\chi$. Another advantage is that no specific approximation regarding the respective intensities of the collisions and of the wave has to be introduced. They can thus be used to validate the results obtained with the adjoint method (see section \ref{sec:adjoint-method}).

In this section, the discussion will focus on the evaluation of $\chi_1$ but it can be easily transposed to $\chi_0$ since the latter is a particular case of the former (with $D_{lh,0}=0$).

The Green function $g_1$ corresponding to Eq.~\ref{eq:kinetic-ec-relax} has been introduced in section \ref{sec:fast-kinetic}. Physically, $g_1(\mbf{u},\mbf{u}',\tau)d\mbf{u}$ is the probability of finding an electron initially at velocity space position $\mbf{u}'$ within element $d\mbf{u}$ at location $\mbf{u}$ after a time $\tau$. It means that the associated steady-state response function $\chi_1(\mbf{u})$, whose definition is given by Eq.~\ref{eq:chi1}, can be evaluated by computing the elemental current carried by each electron of a set whose initial location is $\mbf{u}$ along its relaxation trajectory, as it undergoes the effects of Coulomb collisions and LH wave power, and perform an ensemble average afterwards. Rather than directly solving Eq.~\ref{eq:green-g1}, it is thus possible to resort to a stochastic description of this relaxation process. It can be done by casting this equation into the form
\begin{equation}
\frac{\partial g_1}{\partial\tau}=-\frac{\partial}{\partial\mbf{u}}\cdot\mbf{S}
\label{eq:proba-current}
\end{equation}

Introducing the friction vector $\mbf{F}$ and diffusion tensor $\tensor{D}$ and using Einstein convention for repeated indices, the probability current components are written as
\begin{equation}
S_i=F_ig_1-\frac{\partial}{\partial u_j}D_{ij}g_1
\end{equation}

By identification of Eqs.~\ref{eq:proba-current} and \ref{eq:ccoll-umu}, the diffusion tensor can be written as $\tensor{D}\equiv\tensor{D}_{coll}+\tensor{D}_{lh}$

with
\begin{equation}
\tensor{D}_{coll}=\frac{2}{u^3}
\left( 
\begin{array}{cc}
1 & 0  \\
0 & (Z_i+1)(1-\mu^2)/2
\end{array}
\right)
\end{equation}

and
\begin{equation}
\tensor{D}_{lh}=D_{lh,0}\frac{d_{lh}(u)}{u^2}
\left( 
\begin{array}{cc}
\mu^2 & u\mu(1-\mu^2)  \\
u\mu(1-\mu^2) & (1-\mu^2)^2
\end{array}
\right)
\label{eq:dlh}
\end{equation}

For the force term, $\mbf{F}\equiv\mbf{F}_{coll}+\mbf{F}_{lh}$ with
\begin{equation}
\mbf{F}_{coll}=-\frac{2}{u^3}\Bigg[\bigg(1+\frac{3}{u^2}\bigg)u\uvec{u}+(Z_i+1)\mu\uvec{\mu}\Bigg]
\end{equation}

$\uvec{u}$ and $\uvec{\mu}$ are the unit vectors corresponding to the $u$ and $\mu$ directions.

To compute the drift caused by the wave, it is necessary to bear in mind that when the transformation from one coordinate system to another is not linear, as is the case when the LH quasilinear diffusion coefficient is transformed from $(u_\perp,u_\parallel)$ to $(u,\mu)$ coordinates, the friction term needs to include a contribution from the diffusion in the first coordinate system\cite{Risken1989}. This yields
\begin{equation}
\mbf{F}_{lh}=-\frac{D_{lh,0}}{u^2}\Bigg[\bigg(u\mu d'_{lh}(u)+(1-\mu^2)d_{lh}(u)\bigg)u\uvec{u}+\bigg(u d'_{lh}(u)-3\mu d_{lh}(u)\bigg)(1-\mu^2)\uvec{\mu}\Bigg]
\label{eq:flh}
\end{equation}

with $d'_{lh}\equiv \partial d_{lh}/\partial u_\parallel$.

The Langevin equations describing the electrons trajectories can then be written as\cite{Risken1989}
\begin{equation}
\frac{d\mbf{u}}{d\tau}=\mbf{h}(\mbf{u},\tau)+\tensor{g}(\mbf{u},\tau)\cdot\bds{\xi}(\tau)
\label{eq:langevin-tracks}
\end{equation}

The components of the matrix $\tensor{g}$ are linked to the diffusion tensor $\tensor{D}$ through
\begin{equation}
g_{ij}=(\tensor{D}^{1/2})_{ij}
\end{equation}

where $\tensor{D}^{1/2}$ is obtained by diagonalizing $\tensor{D}$, taking the positive square root of the eigenvalues, and transforming the diagonal matrix back.

In the framework of Stratonovitch calculus, the proper deterministic force components are given by\cite{Risken1989}
\begin{equation}
h_i=F_i-(\tensor{D}^{1/2})_{kj}\frac{\partial}{\partial u_k}(\tensor{D}^{1/2})_{ij}
\end{equation}

The second term of the left-hand side of this equation is a correction to the noise-induced drift. Albeit straightforward, the computation of $\mbf{h}$ and $\tensor{g}$ is tedious, the resulting expressions cumbersome and they will not be presented here.

$\bds{\xi}(\tau)$ is the Gaussian-distributed Langevin force, described by its stochastic properties
\begin{equation}
\langle\xi_i(\tau)\rangle=0\textrm{, and }\langle \xi_i(\tau)\xi_j(\tau')\rangle=2\delta_{ij}\delta(\tau-\tau')
\end{equation}

where $\langle\cdot\rangle$ refers to the average performed over statistical realizations. 

Practically, in the simulations presented in this paper, the collision time is split in numerous timesteps and each electron velocity evolves according to Eq.~\ref{eq:langevin-tracks} until thermalization is attained. It should be emphasized that the integration of a stochastic equation in the case of a multiplicative noise has to be carried out with care to avoid numerical artifacts liable to distort the result. A detailed discussion of this question can be found in Ref.~\onlinecite{GarciaPalacios1998}. Here, both Euler and Heun methods have been implemented and the results have been found to be generally indistinguishable. It is recognized that two main sources of numerical error can alter the solution of stochastic equations\cite{Greiner1988}. Firstly, the statistical error, due to the finite number of realizations, which can be evaluated from standard statistical methods. Secondly, the error induced by the time discretization. To reduce the latter, it is usual to perform several simulations with different timesteps $\Delta\tau$, and extrapolate the result for $\Delta\tau\rightarrow 0$. For the values of $D_{lh,0}$ considered in this section, this error was found to be negligible. 

\begin{figure}[htbp]
\begin{center}
\includegraphics[width=11cm]{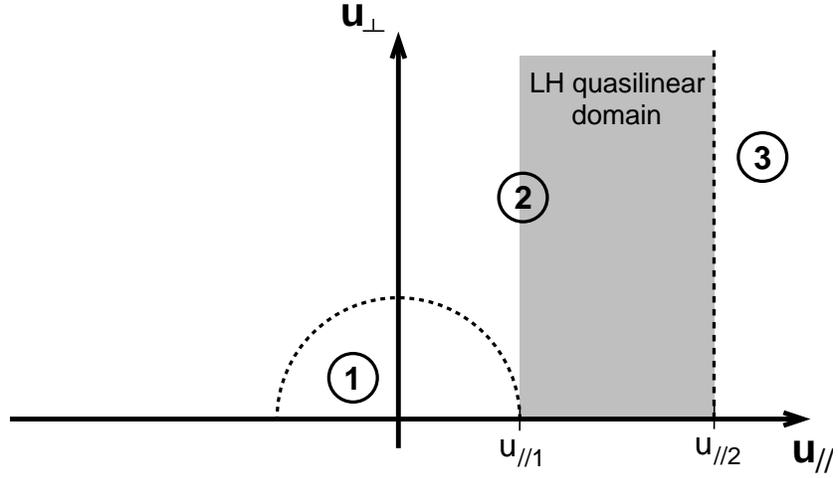}
\end{center}
\caption{Velocity-space configuration in the presence of LH waves. Three domains are distinguished, each corresponding to a different situation with respect to the wave-induced dynamics. Region (1) is such as $u_0<u_{\parallel,1}$, Region (2) encompasses $u_0>u_{\parallel,1}$ and $u_{\parallel 0}<u_{\parallel,2}$, and Region (3) corresponds to $u_{\parallel 0}>u_{\parallel,2}$.}\label{fig:vspace-domains}
\end{figure}

To illustrate the combined effects of Coulomb collisions and LH waves on the electron relaxation, randomly chosen individual relaxation paths can be studied. The velocity-space configuration of the problem appears on Fig.~\ref{fig:vspace-domains}, where three domains (labeled 1-3) are distinguished, corresponding to (1) $u_0<u_{\parallel,1}$, (2) $u_0>u_{\parallel,1}$ and $u_{\parallel 0}<u_{\parallel,2}$, and (3) $u_{\parallel 0}>u_{\parallel,2}$.

\begin{figure}[htbp]
\centering
\includegraphics[width=11cm]{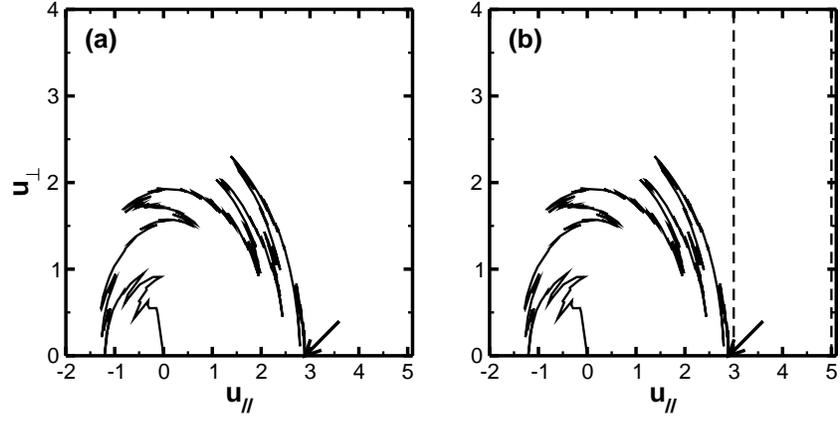}
\caption{Sample relaxation path. (a) Collisions only, (b) Collisions + LH wave. The initial position $(u_{\parallel 0},u_{\perp 0})=(2.9,0)$, is denoted by an arrow. On (b), the dashed lines delimit the LH domain boundaries.}\label{fig:langevin-indv-iv1} 
\end{figure}
\begin{figure}[htbp]
\centering
\includegraphics[width=11cm]{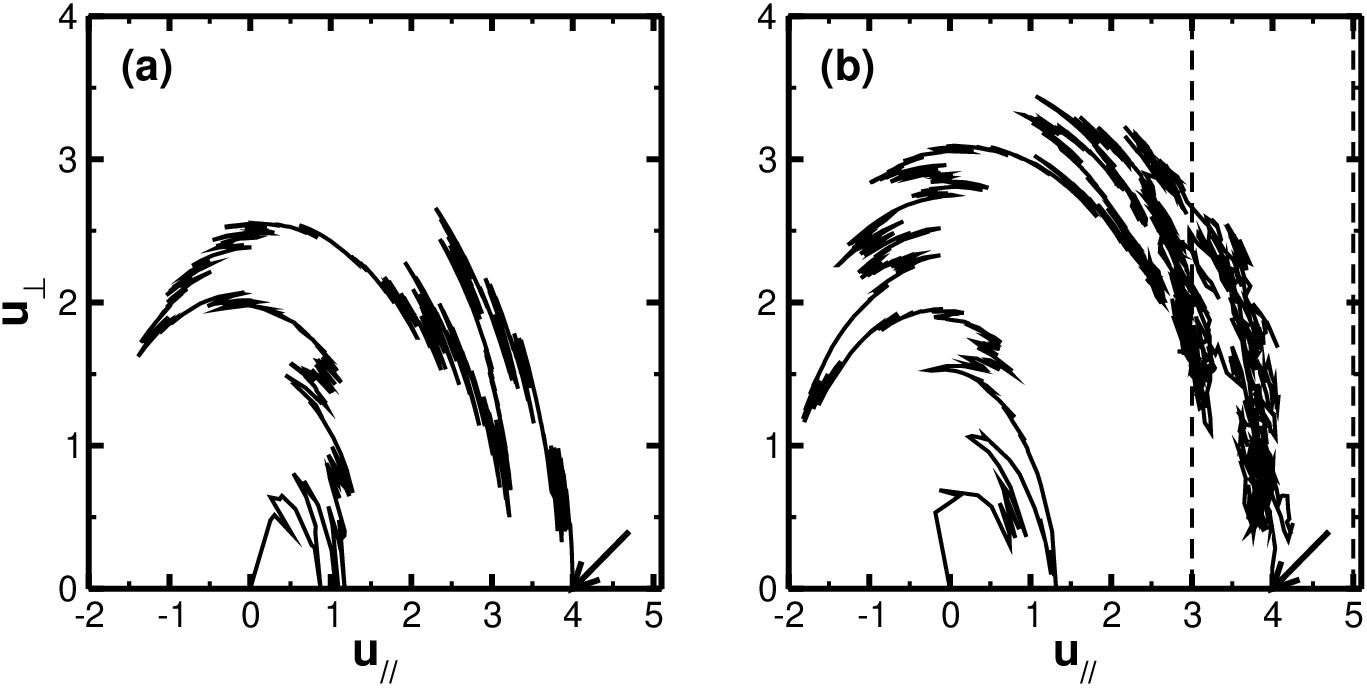}
\caption{Same as Fig.~\ref{fig:langevin-indv-iv1} but for initial position $(u_{\parallel 0},u_{\perp 0})=(4,0)$.}\label{fig:langevin-indv-iv2} 
\end{figure}
\begin{figure}[htbp]
\centering
\includegraphics[width=11cm]{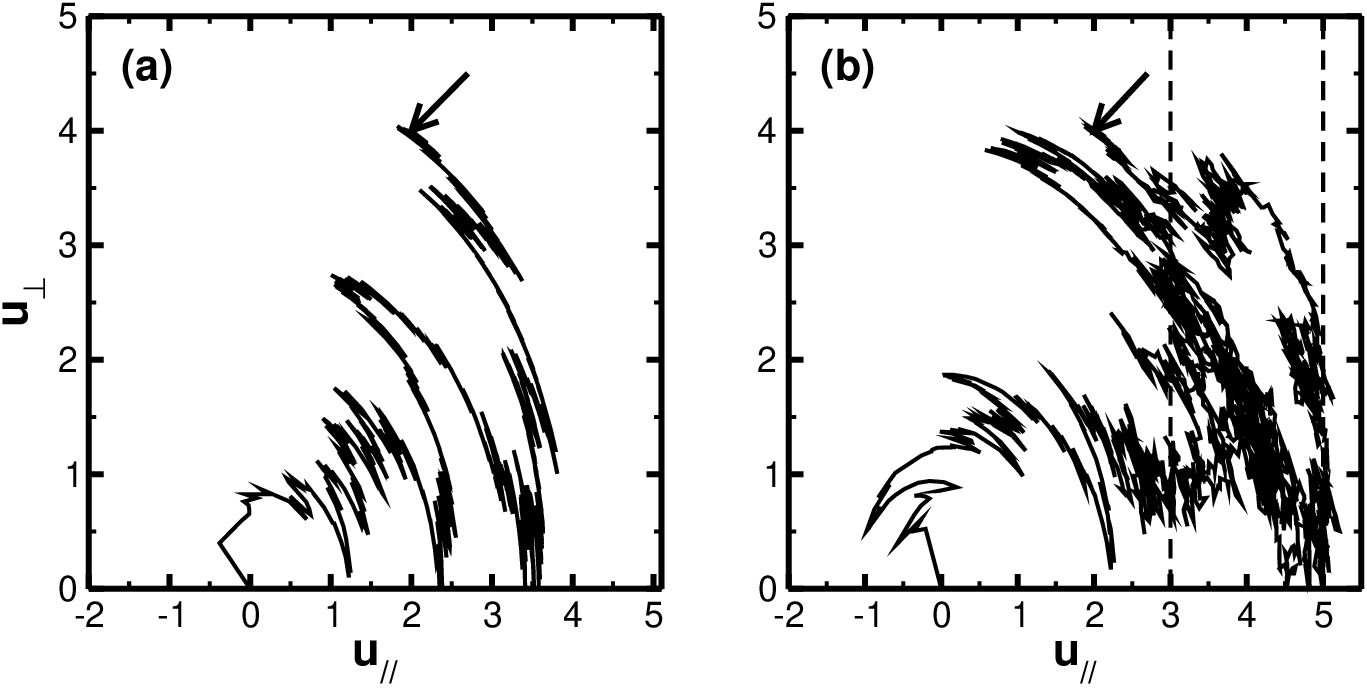}
\caption{Same as Fig.~\ref{fig:langevin-indv-iv1} but for initial position $(u_{\parallel 0},u_{\perp 0})=(2,4)$.}\label{fig:langevin-indv-iv3} 
\end{figure}

Such sample trajectories are shown on Figs.~\ref{fig:langevin-indv-iv1},~\ref{fig:langevin-indv-iv2},~\ref{fig:langevin-indv-iv3} and~\ref{fig:langevin-indv-iv4} which have been produced with the following parameters: $u_{\parallel,1}=3$, $u_{\parallel,2}=5$, $\Delta u_{\parallel,1}=0.5$, $\Delta u_{\parallel,2}=1$, $Z_i=1$ and $D_{lh,0}=0.1$. All these figures are divided in two, the part labelled (a) where only the collisions are included in the calculation, and (b) where the electrons undergo the combined effect of collisions and wave power.





On Fig.~\ref{fig:langevin-indv-iv1}, a trajectory obtained for an electron whose initial velocity lies in region 1 is shown. In such a case and as long as the energy diffusion caused by the collisions is neglected, which is a reasonable approximation, the wave power can not influence the electron relaxation, since the particle never reaches the wave diffusion domain. The relaxation path of an electron starting from the LH quasilinear domain (in region 2) can be seen on Fig.~\ref{fig:langevin-indv-iv2}. In this case, the supplemental parallel diffusion and drift induced by the LH wave clearly lengthen the path, and will be likely to slow the relaxation process, thus enhancing the carried current. Another possibility is for the initial velocity to belong to region 2, but outside the LH domain ($u_{\parallel 0}<u_{\parallel,1}$). A sample trajectory corresponding to this case is shown on Fig.~\ref{fig:langevin-indv-iv3} and interestingly enough, although the particle has an initial parallel velocity such as $u_{\parallel 0}<u_{\parallel,1}$, it experiences the wave influence due to the pitch-angle scattering effect. This is the reason why $u_0$ appears to be as crucial as $u_{\parallel 0}$. Finally, on Fig.~\ref{fig:langevin-indv-iv4}, the relaxation of an electron having $u_{\parallel 0}>u_{\parallel,2}$ is shown (region 3). Although the wave domain is encountered, the net effect on the relaxation length is more complicated, as the upper-velocity boundary of the LH quasilinear coefficient, at least in the model chosen here to describe the wave, induces a drift towards lower velocities (since $\partial d_{lh}/\partial u_\parallel|_{u_{\parallel,2}}<0$) which can accelerate the relaxation with respect to a purely collisional trajectory.

For a quantitative evaluation of the effects of LH waves on the particles thermalization, a statistical analysis has to be performed. This is done by considering many electrons with initial velocity $\mbf{u}_0$, tracking the relaxation trajectories until thermalization and performing the ensemble average, to compute the response function which, in the framework of this stochastic description and according to Eq.~\ref{eq:chi1}, is given by
\begin{equation}
\chi_1(\mbf{u}_0)=\int_0^\infty d\tau'\,\langle u_\parallel\rangle(\tau')
\label{eq:chi1-langevin}
\end{equation}

When the collisions are the only effect taken into account, it is possible to average the Langevin equations analytically and deduce the response function\cite{Fisch1987}. For our purpose, however, this operation is not possible and the computation of the response function has to be performed numerically.

\begin{figure}[htbp]
\centering
\includegraphics[width=11cm]{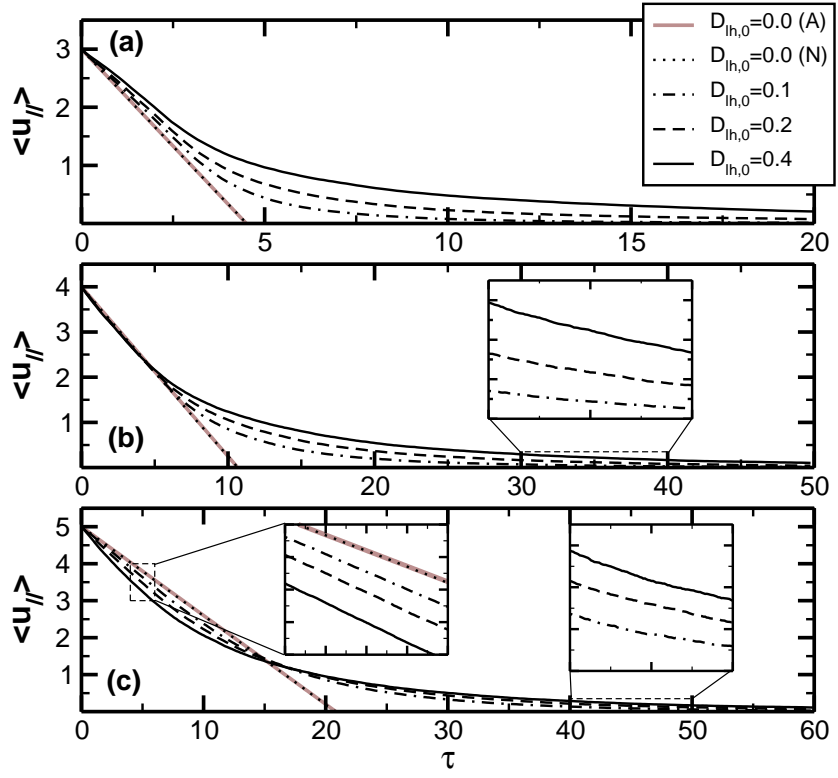}
\caption{$\langle u_\parallel\rangle$ averaged over 20,000 particles as a function of time for $u_{\perp 0}=0$ and (a) $u_{\parallel 0}=3$, (b) $u_{\parallel 0}=4$, and (c) $u_{\parallel 0}=5$. $D_{lh,0}=0$ (dotted line), $D_{lh,0}=0.1$ (dot-dashed line), $D_{lh,0}=0.2$ (dashed line), $D_{lh,0}=0.4$ (solid line). The analytical curve for collisions only appears as a thick grayed line.}\label{fig:langevin-ttraces} 
\end{figure}

Here, we study $\langle u_\parallel\rangle$ as a function of time for various initial velocities and values of $D_{lh,0}$, the average being performed over 20,000 electrons. The result appears on Fig.~\ref{fig:langevin-ttraces}, for initial positions $\mbf{u}_0=(u_{\parallel 0},u_{\perp 0})=(3,0)$, $(4,0)$ and $(6,0)$, and normalized LH diffusion coefficients $D_{lh,0}=0$ (collisions only), $D_{lh,0}=0.1$, $D_{lh,0}=0.2$, $D_{lh,0}=0.4$. The collisional Fisch-Boozer solution, given (for $Z_i=1$) by $\langle u_\parallel\rangle(\tau)=u_{\parallel 0}(1-6\tau/u_0^3)$ also appears but is perfectly superimposed with its numerical counterpart and is thus barely visible.


It can be observed that for $(u_{\parallel 0},u_{\perp 0})=(3,0)$ or $(4,0)$, the wave clearly delays the thermalization. In the third case $(u_{\parallel 0}=6)$, the effect of the drift induced at the high-velocity boundary appears clearly, since the electrons begin by experiencing a faster decrease in parallel velocity, on average. However, the energy range of the electrons is largely spread by the wave and a significant proportion has not yet thermalized, well after the purely collisional relaxation is over. The response function is determined by the balance between these two effects and its features shall be presented in the next section.

\section{Response function of a LHCD plasma}
\label{sec:response-function-lhcd}
In section \ref{sec:langevin-equations}, a numerical method has been presented and employed to perform a basic analysis of the velocity space structure when the presence of LH power influences the dynamics underlying the current drive process. Although it can be used to compute the response function, the perturbation method presented in section \ref{sec:adjoint-method} is more economical in terms of computational resources and is thus more adapted to a systematic study of the response function properties, which is the goal of the present section.

\begin{figure}[htbp]
\centering
\includegraphics[width=11cm]{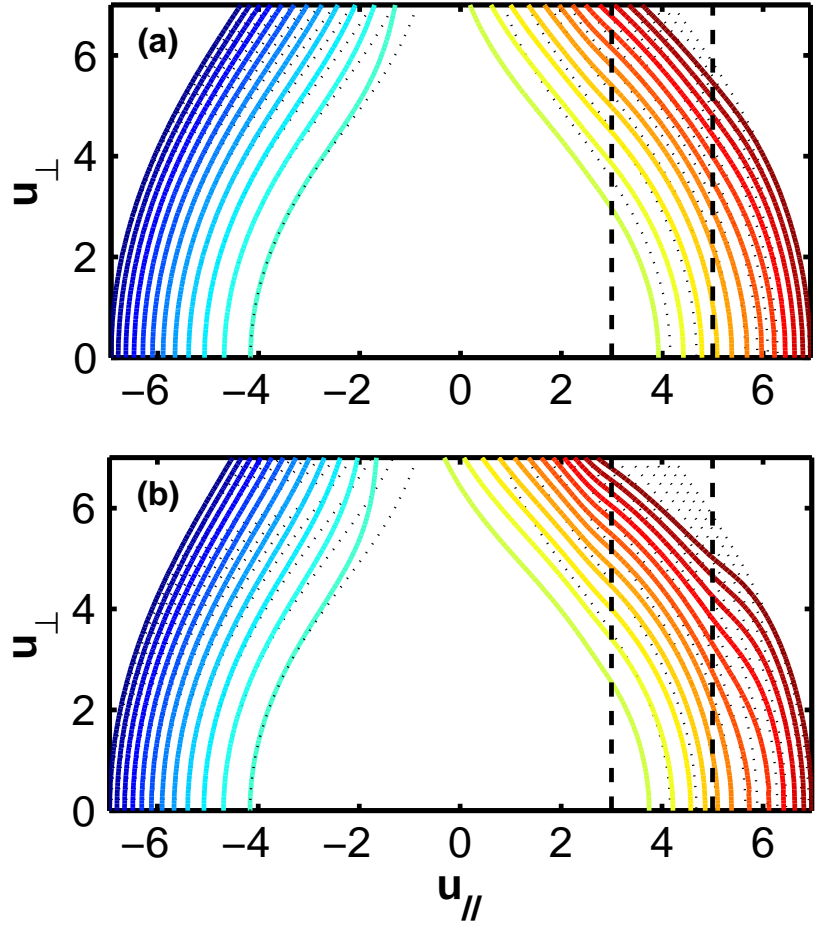}
\caption{Iso-contours of the response function $\chi_0+\delta\chi$ of a plasma with LH waves such as $u_{\parallel,1}=3$, $u_{\parallel,2}=5$, $\Delta u_{\parallel,1}=0.5$, $\Delta u_{\parallel,2}=1$ and (a) $D_{lh,0}=0.1$, (b) $D_{lh,0}=0.2$. The dashed vertical lines delimit the LH domain boundaries and the dotted contours represent the collisional response function $\chi_0$. Contours start at $|\chi|=25$ and are equally spaced with $|\Delta\chi|=15$.}\label{fig:chi-2dcontours-cl} 
\end{figure}

In what follows, the same parameters as in section \ref{sec:langevin-equations} are considered for the Lower Hybrid quasilinear domain. On Fig.~\ref{fig:chi-2dcontours-cl}, some level curves the total response function $\chi_0+\delta\chi$ are represented in $(u_{\parallel},u_{\perp})$ space for $D_{lh,0}=0.1$ (a) and $D_{lh,0}=0.2$ (b). For comparison, the corresponding contours of the Fisch-Boozer response function $\chi_0$ appear as dashed lines.

\begin{figure}[htbp]
\centering
\includegraphics[width=11cm]{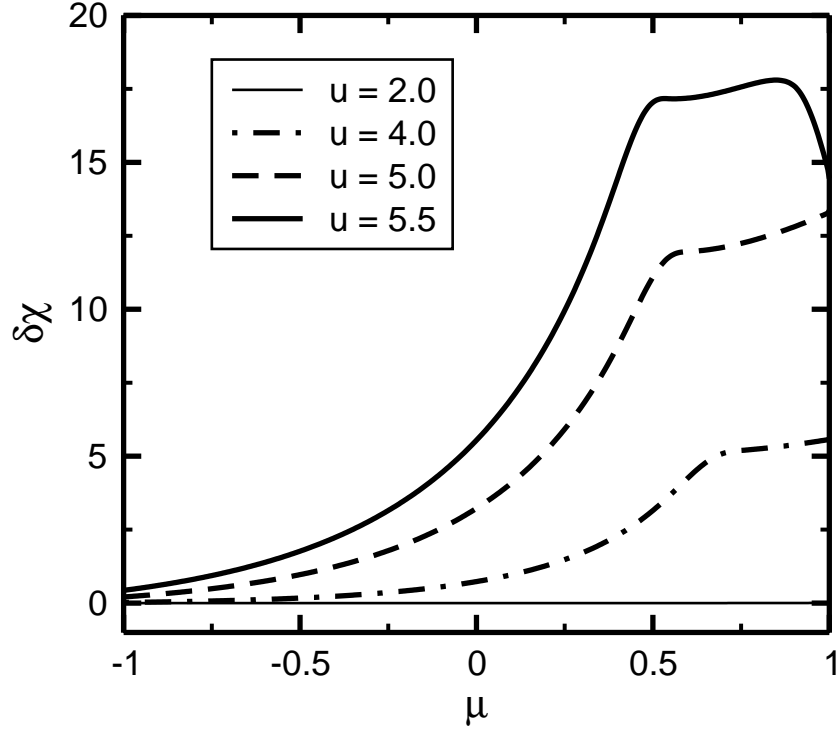}
\caption{$\delta\chi$ as a function of the pitch-angle cosine $\mu$ for $u=2$ (thin solid), $4$ (dot-dashed), $5$ (dashed) and $5.5$ (thick solid). The plasma and LH wave parameters are the same as on Fig.~\ref{fig:chi-2dcontours-cl}(a).}\label{fig:dchi-mu} 
\end{figure}

Several observations can be made about this figure. Firstly, the overall modification of the response function is rather moderate, which is consistent with the approximation of the adjoint calculation, which requires the collisions to dominate the electron relaxation. Secondly, the response function is modified mainly in the LH quasilinear domain, but not only and it can be seen to extend to all velocities such as $u>u_{\parallel,1}$ as well as beyond $u_{\parallel}=u_{\parallel,2}$, which is consistent with the conclusions drawn in section \ref{sec:langevin-equations}. This behavior is clearly visible on Fig.~\ref{fig:dchi-mu}, where the response function perturbation $\delta\chi$ is shown as a function of $\mu$ for various values of $u$, and $D_{lh,0}=0.1$.

\begin{figure}[htbp]
\centering
\includegraphics[width=11cm]{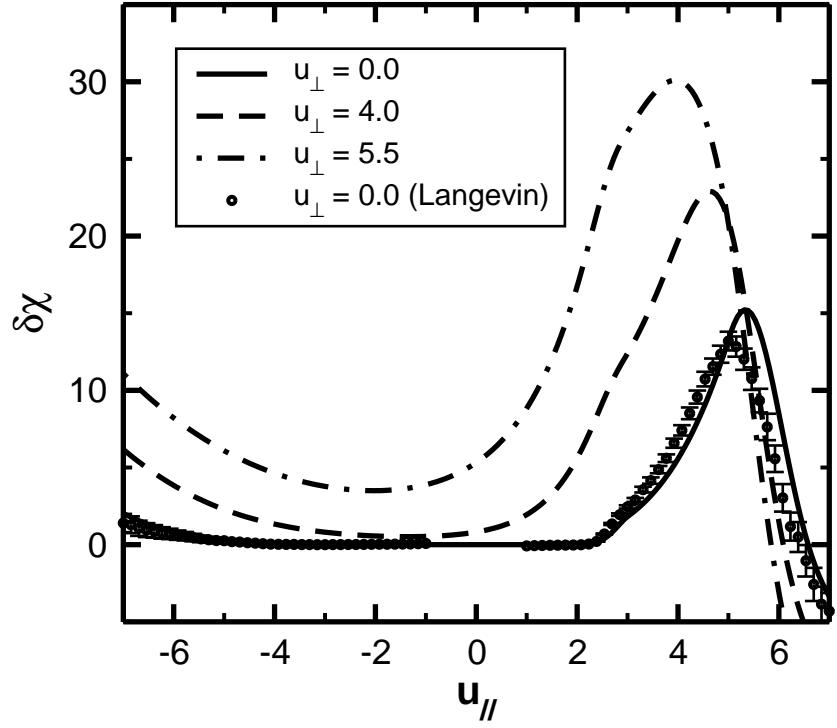}
\caption{$\delta\chi$ as a function of normalized parallel velocity $u_\parallel$ for $u_\perp=0$ (solid), $4$ (dashed), and $5.5$ (dot-dashed). Also shown is the response function obtained by numerical solution of the Langevin equations for $u_\perp=0$ (crosses).}\label{fig:dchi-upar} 
\end{figure}

This figure shows the strong asymmetrical shape of $\delta\chi$. As a result, the total response function $\chi_1$ is largely enhanced in the $u_\parallel>0$ region of velocity space, under the influence of the LH wave. As predicted for $u<u_{\parallel,1}$, we obtain $\delta\chi=0$, in other words $\chi_1$ reduces to the Fisch-Boozer response function. The pitch-angle scattering effect can cause $\delta\chi(u,\mu)$ to be non-zero even for $u_\parallel<u_{\parallel,1}$ and particularly for $u_{\parallel}<0$. Another observation is that $\delta\chi$ falls off rapidly for $u>u_{\parallel,2}$ and can even become negative. This effect is best viewed when $\delta\chi$ is represented as a function of $u_{\parallel}$, for various values of $u_{\perp}$, as shown on Fig.~\ref{fig:dchi-upar}. For completeness, the result from the Langevin equations computation also appears for $u_\perp=0$, the error bars being deduced from the estimated statistical error.

\begin{figure}[htbp]
\centering
\includegraphics[width=11cm]{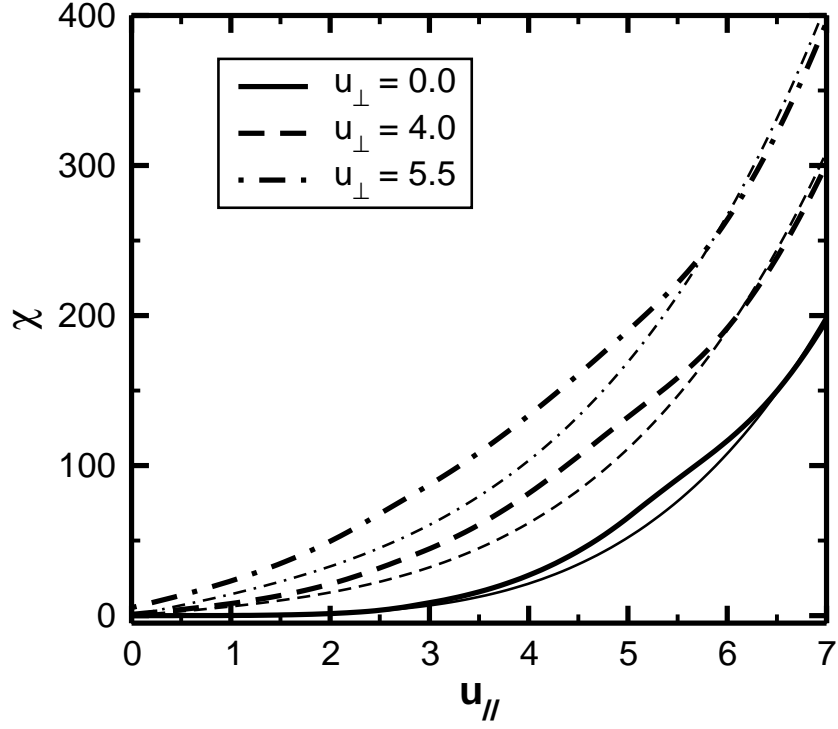}
\caption{Comparison of the Fisch-Boozer response function $\chi_0$ (thin lines) and the response function including LH wave effects $\chi_0+\delta\chi$ (thick lines) for $D_{lh,0}=0.1$ for $u_\perp=0$ (solid), $4$ (dashed) and $5.5$ (dot-dashed).}\label{fig:chi-upar}
\end{figure}

One can notice that for $u\gtrsim 6$, $\delta\chi$ is negative, which would indicate a deleterious effect of the LH wave on the EC-driven current when the latter is carried by electrons excited in this region. Although supported by the numerical solution of the Langevin equations, this conclusion must, however, be tempered by several considerations: \emph{(i)} In this region of velocity space, $\delta\chi$ is sensitive to the value chosen for $\Delta u_{\parallel,2}$ which is not readily available, \emph{(ii)} EC wave absorption at velocities significantly above the upper bound of the LH quasilinear domain is difficult and would most likely be impossible in the absence of LH wave. The concept of EC current drive improvement or degradation is thus of little sense, \emph{(iii)} The Fisch-Boozer response function is proportional to $u^4$, which makes the LH-induced modification rather weak for large values of $u_\parallel$. The latter point is supported by the result of Fig.~\ref{fig:chi-upar}, where $\chi_0$ and $\chi_0+\delta\chi$ are shown as a function of $u_\parallel$ for the same parameters as Fig.~\ref{fig:dchi-upar}.


This figure shows that beyond the upper boundary of the LH domain, the effect is indeed small. Physically, this simply means that no EC efficiency improvement takes place for $u_{\parallel}\gtrsim u_{\parallel,2}$, aside from the fact that the very presence of a LH plateau is responsible for the EC wave absorption at this location\cite{Fidone1984}.

\begin{figure}[htbp]
\centering
\includegraphics[width=11cm]{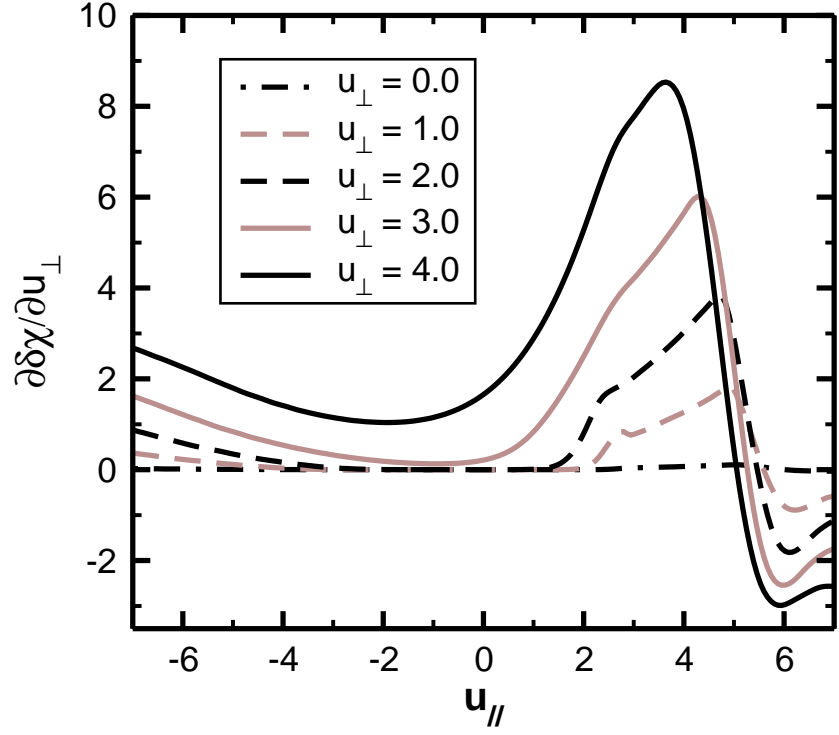}
\caption{$\partial\delta\chi/\partial u_\perp$ as a function of $u_\parallel$ for $u_\perp=0$ (dot-dashed), $1$ (dashed grayed), $2$ (dashed), $3$ (solid grayed) and $4$ (solid).}\label{fig:ddchiduper-upar} 
\end{figure}

More important than the  response function itself, as far as the RF current drive is concerned, is its velocity-space gradient, as is apparent from Eq.~\ref{eq:def-j1}. For EC waves, the differentiation is to be performed along $u_\perp$. On Fig.~\ref{fig:ddchiduper-upar}, the quantity $\delta\chi'\equiv\partial\delta\chi/\partial u_\perp$ is shown versus $u_\parallel$ for various values of $u_\perp$.


Whenever $u_\parallel<u_{\parallel,2}$, $\delta\chi'$ appears to be positive, which implies a favorable contribution of the LH wave to the EC current. Moreover, even for moderate values of $u_\perp$, where electrons are most easily driven by EC waves, $\delta\chi'$ can be fairly large, provided an appropriate range of parallel velocities is selected, which is possible through the use of suitable launching angles\cite{Erckmann1994}. For $u_\parallel\gtrsim u_{\parallel,2}$, $\delta\chi'$ can have a negative value, although as stated above, this feature should be pondered cautiously. This figure confirms that the LH wave has an overall beneficial effect on the current driven EC wave, and that a synergy between the two waves can be expected, especially when the latter are excited in the vicinity of $u_{\parallel,ec}\lesssim u_{\parallel,2}$. Note that this parameter is generally simply determined by the LH wave accessibility condition\cite{Bonoli1986}, and is therefore readily available from experimental measurements of the major plasma parameters.

\section{ECCD efficiency in the presence of LH waves}
\label{sec:lhec-efficiency}
The features of the distribution function in the presence of LH waves presented in the previous section have consequences in terms of ECCD efficiency. From Eq.~\ref{eq:normcurrent}, the total current appears as $j=j_0+j_1$ where $j_0$ (Eq.~\ref{eq:def-j0}) is driven by the LH wave, and is implicitly assumed to be unaffected by the presence of the EC wave, in the present model. In section \ref{sec:adjoint-method}, the response function of the plasma in the presence of LH waves was linearized according to $\chi_1=\chi_0+\delta\chi$, which implies that $j_1$, given by Eq.~\ref{eq:def-j1} can be cast into the form $j_1=j_{ec}+\delta j$, with
\begin{equation}
j_{ec}\equiv\int d\mbf{u}\,\mbf{S}_{ec}\cdot\frac{\partial\chi_0}{\partial\mbf{u}}
\end{equation}

and
\begin{equation}
\delta j\equiv\int d\mbf{u}\,\mbf{S}_{ec}\cdot\frac{\partial\delta\chi}{\partial\mbf{u}}
\end{equation}

$j_{ec}$ is the EC current obtained when the cross-effect of both waves is not accounted for and $\delta j$ is a supplemental current, which qualifies for the denomination \emph{synergy current} (or anti-synergy, in the event that $\delta j$ and $j_{ec}$ have opposite signs). This clear distinction stems from the linearization introduced in the computation of the response function and allows a straightforward separation of the contribution of each process, which is generally the key difficulty encountered when trying to characterize a synergy effect.

Defining $j_w$ the amount of current generated by wave power $p_w$, the steady-state current drive efficiency can be expressed as\cite{Fisch1987}
\begin{equation}
\eta_w\equiv\frac{j_w}{p_w}=\frac{\ds{\int d\mbf{u}\,\mbf{S}_w\cdot\frac{\partial\chi}{\partial\mbf{u}}}}{\ds{\int d\mbf{u}\,\mbf{S}_w\cdot\frac{\partial}{\partial\mbf{u}}\bigg(\frac{u^2}{2}\bigg)}}
\end{equation}

where $\chi$ is the associated response function and $\mbf{S}_w$ the quasilinear flux. Uf the interaction is supposed to be localized in velocity-space, as is generally the case with EC waves and, although to a lesser extent, with LH waves also, the efficiency can be approximated to give
\begin{equation}
\eta_w\approx\frac{\ds{\mbf{S}_w\cdot\frac{\partial\chi}{\partial\mbf{u}}}}{\ds{\mbf{S}_w\cdot\frac{\partial}{\partial\mbf{u}}\bigg(\frac{u^2}{2}\bigg)}}
\label{eq:local-efficiency}
\end{equation}

Evaluating this expression only requires the direction of $\mbf{S}_w$, which unlike its magnitude, is only weakly dependent on the precise shape of the distribution function and is well-known\cite{Fisch1987} ($\mbf{S}_{ec}\propto\uvec{\perp}$ and $\mbf{S}_{lh}\propto\uvec{\parallel}$). 

\begin{figure}[htbp]
\centering
\includegraphics[width=11cm]{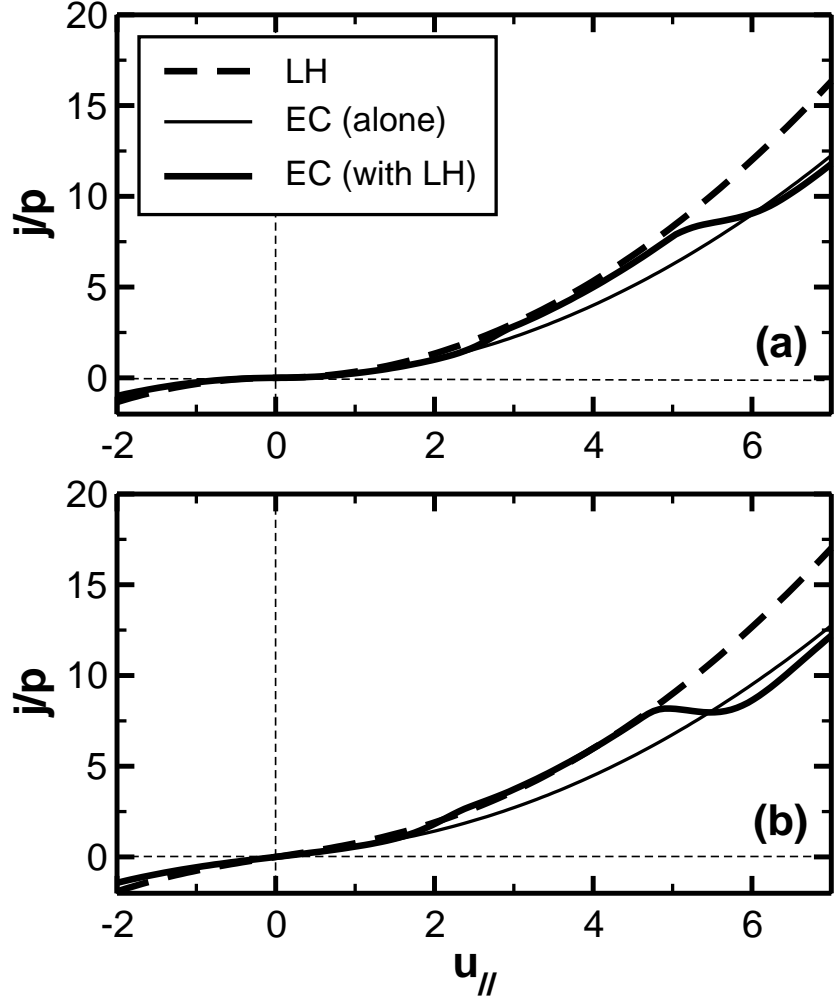}
\caption{Current drive efficiency as a function of $u_\parallel$ for (a) $u_\perp=0$ and (b) $u_\perp=2$ for LH waves (dashed line) for EC waves alone (thin solid line) and for EC waves when the synergy effect with LH waves is taken into account with $D_{lh,0}=0.1$ (thick solid line).}\label{fig:eff-upar-uper12}
\end{figure}

Here, the efficiencies of both waves are computed using Eq.~\ref{eq:local-efficiency} with the same parameters as in Section \ref{sec:adjoint-method}: $D_{lh,0}=0.1$, $u_{\parallel,1}=3$, $u_{\parallel,2}=5$, $\Delta u_{\parallel,1}=0.5$ and $\Delta u_{\parallel,2}=1$. On Fig.~\ref{fig:eff-upar-uper12}, these efficiencies are shown as functions of $u_\parallel$ for $Z_i=1$, $u_\perp=0$ and $u_\perp=2$


The 4:3 ratio between the LH and EC efficiencies is recovered, in the absence of a cross-effect\cite{Fisch1987}. If this effect is included, the corrected EC efficiency (i.e. $(j_{ec}+\delta j)/p_{ec}$) is significantly enhanced in the region of velocity space corresponding to the LH superthermal plateau. For the chosen value of $D_{lh,0}$, it becomes comparable to the LH efficiency. A slight anti-synergy effect is observed for $u_{\parallel}$ above $u_{\parallel,2}$ but this point has been extensively addressed in Section \ref{sec:adjoint-method}. The EC efficiency enhancement increases with $u_\perp$ and can exceed the LH efficiency on a significant range of parallel velocities. Moreover, for some parameters, the counter-current drive amount can be slightly lowered by the synergy effect, which can further enhance the current driven in a EC downshift scheme\cite{Erckmann1994}, where the wave interacts with electrons having $u_\parallel<0$ as well as $u_\parallel>0$.

To characterize the ECCD efficiency enhancement, following the definition introduced in Ref.~\onlinecite{Dumont2000}, the improvement factor $F_{syn}\equiv (j_{lh+ec}-j_{lh})/j_{ec}$ is studied. According to the linearization introduced in this model, it simplifies to give 
\begin{equation}
F_{syn}=1+\frac{\delta j}{j_{ec}}=1+\frac{\partial\delta\chi/\partial u_\perp}{\partial\chi_0/\partial u_\perp}
\label{eq:fsyn}
\end{equation}

\begin{figure}[htbp]
\centering
\includegraphics[width=11cm]{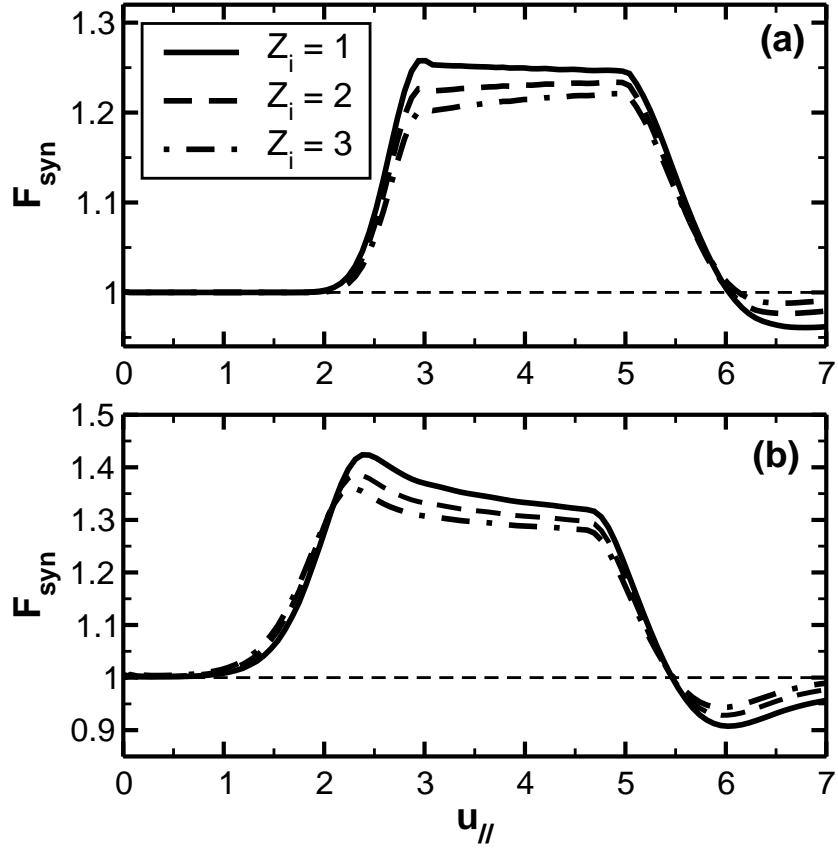}
\caption{Synergy factor in the presence of LH waves for $D_{lh,0}=0.1$ and various values of the plasma ion charge: $Z_i=1$ (solid), $Z_i=2$ (dashed), and $Z_i=3$ (dot-dashed). (a) $u_\perp=0$, (b) $u_\perp=2$.}\label{fig:zeff-upar}
\end{figure}

The same parameters as above are used, but the ion charge $Z_i$ is varied from $1$ to $3$. The variation of the synergy factor as a function of $u_\parallel$ is shown on Fig.~\ref{fig:zeff-upar}.


It is seen that the interaction with electrons whose parallel velocity lies in the vicinity of the LH quasilinear domain can be very beneficial for the EC current drive. For the chosen value of $D_{lh,0}$, an improvement of the efficiency as high as 40\% can be obtained. The anti-synergy effect underlined above proves to be marginal. Another observation is that even though the plasma ion charge increase is known to be detrimental to superthermal electrons-based current drive schemes, it has only a minor influence on the synergy mechanism itself. Finally, a particularly noticeable feature is that the improvement factor is found to be weakly dependent on the particular velocity space location under consideration.

Hitherto, the presented simulations have all been performed for $D_{lh,0}=0.1$, in order to ensure the validity of the perturbation technique employed to derive the response function from the adjoint equation. The Langevin equations formalism, introduced in section \ref{sec:langevin-equations}, is more demanding from a computational point of view. Still, it has the advantage of offering more flexibility than a Fokker-Planck treatment and allows to study how the EC Current Drive efficiency depends on $D_{lh,0}$, since it is derived without any assumption regarding its particular value. 

This study is performed by solving the Langevin equations (Eq.~\ref{eq:langevin-tracks}) for 20,000 electrons at each initial velocity. For increasing values of $D_{lh,0}$, the timestep $\Delta\tau$ is decreased so as to remain small with respect to the variations of $\tensor{D}_{lh}$ (see Eq.~\ref{eq:dlh}) and $\mbf{F}_{lh}$ (Eq.~\ref{eq:flh}). To ensure convergence, however, the computation is performed with several values of $\Delta\tau$, and the resulting averaged quantities are extrapolated to $\Delta\tau\rightarrow 0$. For large values of $D_{lh,0}$, this procedure proves to be necessary to avoid the inherent bias induced by the time discretization, which adds to the purely statistical error\cite{Greiner1988}.

The same plasma and LH wave parameters as above are chosen, with $Z_i=1$. The simulations are performed for velocity space location $(u_\parallel,u_\perp)=(4,1)$ and the quantity under study is $\delta\chi\equiv \chi_1-\chi_0$ where $\chi_1$ is obtained from the Langevin equations (Eq.~\ref{eq:chi1-langevin}) and $\chi_0$ is given by Eq.~\ref{eq:chi-fischboozer}. As discussed in section \ref{sec:lhec-efficiency}, the most relevant quantity, as far as the ECCD efficiency is concerned, is $\delta\chi'=\partial\delta\chi/\partial u_\perp$. To obtain it, simulations are performed for several values of $u_\perp$ and the derivative is obtained numerically.

\begin{figure}[htbp]
\centering
\includegraphics[width=11cm]{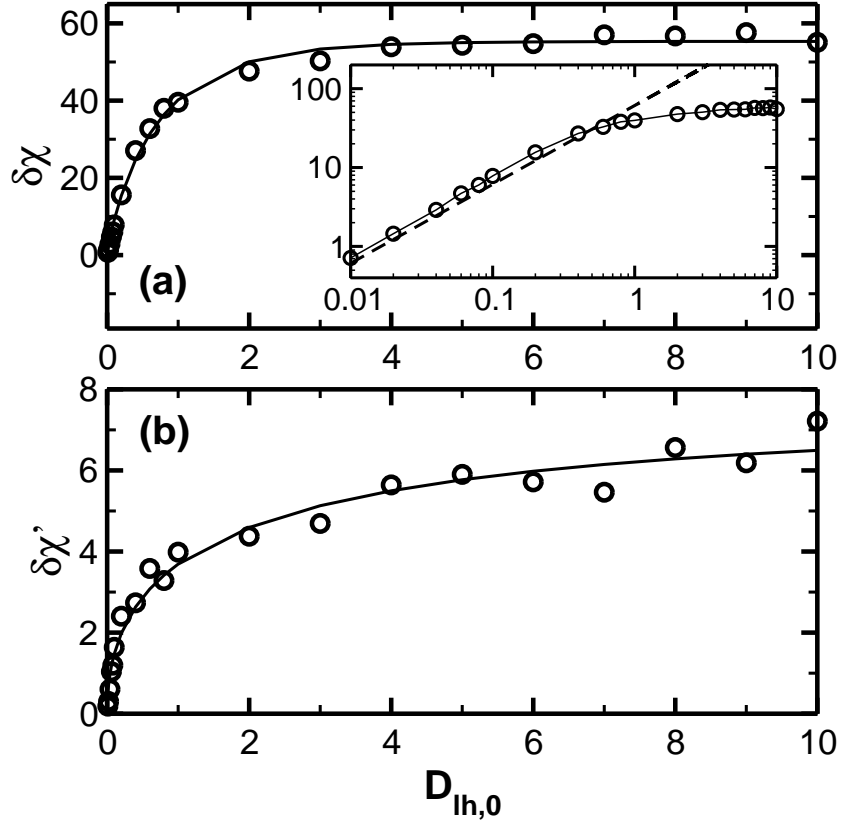}
\caption{(a) $\delta\chi$ and (b) $\partial\delta\chi/\partial u_\perp$ versus normalized LH quasilinear diffusion coefficient $D_{lh,0}$ for $u_\parallel=4$ and $u_\perp=1$. The solid lines are fitting curves and on (a), the result of the adjoint method is shown as a dashed line in the inset containing $\delta\chi$ plotted on logarithm axes.}\label{fig:deltachi-saturation}
\end{figure}

On Fig.~\ref{fig:deltachi-saturation}, $\delta\chi$ and $\delta\chi'$ are plotted versus $D_{lh,0}$, with associated fitting curves. For comparison, the adjoint solution for $\delta\chi$  appears also.


The adjoint solution appears to give a fair result for $D_{lh,0}\lesssim 1$, in accordance with the validity range of the associated method. Its linear dependence on $D_{lh,0}$ (see Eq.~\ref{eq:deltachi}), however, leads to an overestimate of the response function, as $\delta\chi$ is found to level off when $D_{lh,0}$ is increased. To extrapolate the results as $D_{lh,0}$ tends to infinity, i.e. in a perfectly saturated situation, the following fitting function is employed for $\delta\chi'$
\begin{equation}
\delta\chi'=a_0\cdot (1-\exp(-a_1 D_{lh,0}^{a_2}))
\end{equation}

A least square fit leads to $a_0\approx 7.3$, $a_1\approx 0.7$ and $a_2\approx 0.5$. This gives the extrapolated value $\lim_{D_{lh,0}\rightarrow\infty}\delta\chi'\approx 7.3$, or when used in Eq.~\ref{eq:fsyn}, $F_{syn}\approx 2.8$. This means that for the parameters considered here, the ECCD efficiency is nearly tripled when compared to its ``standard'' value, i.e. the value obtained in the absence of Lower Hybrid waves.

\section{Conclusions}
\label{sec:conclusions}
Owing to the complexity of a full kinetic treatment of the current drive problem in tokamaks, added to the difficulty of separating the contributions from various physical processes, the existence of a synergy between LH and EC waves has often been disputed. In this paper, a different approach has been employed to address this question from a theoretical standpoint, when the dominant source of deformation of the distribution function is LH power. Two complementary methods have been employed: a perturbation solution of the adjoint equation allows a fast derivation of the response function including both collisional and LH effects in the limit where the former dominate. On the other hand, solving the Langevin equations allows to overcome this restriction and in addition to the computation of the response function, provides clear insight in the dynamics underlying the process. However, they imply a less straightforward mathematical treatment and higher computational requirements.

By application of these two formalisms, it has been shown that a synergy was indeed possible between the two waves, provided the EC parameters are chosen to drive electrons within or close to the LH quasilinear domain. Even for moderate values of the LH quasilinear diffusion coefficient, a significant improvement of the ECCD efficiency has been obtained. Moreover, for sufficiently high values of this coefficient, i.e. when quasilinear saturation is reached, the improvement factor appears to be nearly constant. Also, in this study, the efficiency enhancement has been found to exhibit only a mild dependence on the particular velocity space location. These trends suggest that the synergy mechanism is fairly robust and should manifest itself provided the EC waves are launched using a set of parameters compatible with the LH quasilinear domain properties. 

In present experiments, the characterization of such an enhancement of EC current drive efficiency can be a daunting task. This is mostly due to the fact that, as in numerical simulations, a lot of phenomena are involved in the process and are difficult to separate. It is nonetheless possible to envision experimental scenarios aimed at studying the LH-EC synergy. One such scenario could be creating a fully non-inductive plasma by relying on the LHCD system. After a delay equivalent to several resistive times, injecting two EC beams with opposite parallel spectra - i.e. opposite toroidal angle, to lowest order - should result in zero net EC current in the absence of a cross-effect. The measurement of an additional amount of current would then be the signature of a synergy between the two waves.

\acknowledgments
One of the authors (RJD) wishes to express his appreciation for encouraging comments by M. Brambilla and G. Leclert.

\bibliography{synergy-lhec}

\begin{thebibliography}{28}
\expandafter\ifx\csname natexlab\endcsname\relax\def\natexlab#1{#1}\fi
\expandafter\ifx\csname bibnamefont\endcsname\relax
  \def\bibnamefont#1{#1}\fi
\expandafter\ifx\csname bibfnamefont\endcsname\relax
  \def\bibfnamefont#1{#1}\fi
\expandafter\ifx\csname citenamefont\endcsname\relax
  \def\citenamefont#1{#1}\fi
\expandafter\ifx\csname url\endcsname\relax
  \def\url#1{\texttt{#1}}\fi
\expandafter\ifx\csname urlprefix\endcsname\relax\def\urlprefix{URL }\fi
\providecommand{\bibinfo}[2]{#2}
\providecommand{\eprint}[2][]{\url{#2}}

\bibitem[{\citenamefont{Fisch}(1987)}]{Fisch1987}
\bibinfo{author}{\bibfnamefont{N.~J.} \bibnamefont{Fisch}},
  \bibinfo{journal}{Rev. Mod. Physics} \textbf{\bibinfo{volume}{59}},
  \bibinfo{pages}{175} (\bibinfo{year}{1987}).

\bibitem[{\citenamefont{Taylor}(1997)}]{Taylor1997}
\bibinfo{author}{\bibfnamefont{T.~S.} \bibnamefont{Taylor}},
  \bibinfo{journal}{Plasma Phys. Contr. Fusion} \textbf{\bibinfo{volume}{39}},
  \bibinfo{pages}{B47} (\bibinfo{year}{1997}).

\bibitem[{\citenamefont{Litaudon et~al.}(1996)\citenamefont{Litaudon,
  Arslanbekov, Hoang, Joffrin, Kazarian-Vibert, Moreau, Peysson, Bibet,
  Froissard, Goniche et~al.}}]{Litaudon1996}
\bibinfo{author}{\bibfnamefont{X.}~\bibnamefont{Litaudon}},
  \bibinfo{author}{\bibfnamefont{R.}~\bibnamefont{Arslanbekov}},
  \bibinfo{author}{\bibfnamefont{G.~T.} \bibnamefont{Hoang}},
  \bibinfo{author}{\bibfnamefont{E.}~\bibnamefont{Joffrin}},
  \bibinfo{author}{\bibfnamefont{F.}~\bibnamefont{Kazarian-Vibert}},
  \bibinfo{author}{\bibfnamefont{D.}~\bibnamefont{Moreau}},
  \bibinfo{author}{\bibfnamefont{Y.}~\bibnamefont{Peysson}},
  \bibinfo{author}{\bibfnamefont{P.}~\bibnamefont{Bibet}},
  \bibinfo{author}{\bibfnamefont{P.}~\bibnamefont{Froissard}},
  \bibinfo{author}{\bibfnamefont{M.}~\bibnamefont{Goniche}},
  \bibnamefont{et~al.}, \bibinfo{journal}{Plasma Phys. Control. Fusion}
  \textbf{\bibinfo{volume}{38}}, \bibinfo{pages}{1603} (\bibinfo{year}{1996}).

\bibitem[{\citenamefont{Barbato}(1998)}]{Barbato1998}
\bibinfo{author}{\bibfnamefont{E.}~\bibnamefont{Barbato}},
  \bibinfo{journal}{Plasma Phys. Control. Fusion}
  \textbf{\bibinfo{volume}{40}}, \bibinfo{pages}{A63} (\bibinfo{year}{1998}).

\bibitem[{\citenamefont{{Y. Peysson and the Tore Supra
  Team}}(2000)}]{Peysson2000}
\bibinfo{author}{\bibnamefont{{Y. Peysson and the Tore Supra Team}}},
  \bibinfo{journal}{Plasma Phys. Control. Fusion}
  \textbf{\bibinfo{volume}{42}}, \bibinfo{pages}{B87} (\bibinfo{year}{2000}).

\bibitem[{\citenamefont{Erckmann and Gasparino}(1994)}]{Erckmann1994}
\bibinfo{author}{\bibfnamefont{V.}~\bibnamefont{Erckmann}} \bibnamefont{and}
  \bibinfo{author}{\bibfnamefont{U.}~\bibnamefont{Gasparino}},
  \bibinfo{journal}{Plasma Phys. Control. Fusion}
  \textbf{\bibinfo{volume}{36}}, \bibinfo{pages}{1869} (\bibinfo{year}{1994}).

\bibitem[{\citenamefont{Fidone et~al.}(1984)\citenamefont{Fidone, Giruzzi,
  Granata, and Meyer}}]{Fidone1984}
\bibinfo{author}{\bibfnamefont{I.}~\bibnamefont{Fidone}},
  \bibinfo{author}{\bibfnamefont{G.}~\bibnamefont{Giruzzi}},
  \bibinfo{author}{\bibfnamefont{G.}~\bibnamefont{Granata}}, \bibnamefont{and}
  \bibinfo{author}{\bibfnamefont{R.~L.} \bibnamefont{Meyer}},
  \bibinfo{journal}{Phys. Fluids} \textbf{\bibinfo{volume}{27}},
  \bibinfo{pages}{2468} (\bibinfo{year}{1984}).

\bibitem[{\citenamefont{Fidone et~al.}(1987)\citenamefont{Fidone, Giruzzi,
  Krivenski, Mazzucato, and Ziebell}}]{Fidone1987}
\bibinfo{author}{\bibfnamefont{I.}~\bibnamefont{Fidone}},
  \bibinfo{author}{\bibfnamefont{G.}~\bibnamefont{Giruzzi}},
  \bibinfo{author}{\bibfnamefont{V.}~\bibnamefont{Krivenski}},
  \bibinfo{author}{\bibfnamefont{E.}~\bibnamefont{Mazzucato}},
  \bibnamefont{and} \bibinfo{author}{\bibfnamefont{L.~F.}
  \bibnamefont{Ziebell}}, \bibinfo{journal}{Nucl. Fusion}
  \textbf{\bibinfo{volume}{27}}, \bibinfo{pages}{579} (\bibinfo{year}{1987}).

\bibitem[{\citenamefont{Farina et~al.}(1988)\citenamefont{Farina, Lontano, and
  Pozzoli}}]{Farina1988}
\bibinfo{author}{\bibfnamefont{D.}~\bibnamefont{Farina}},
  \bibinfo{author}{\bibfnamefont{M.}~\bibnamefont{Lontano}}, \bibnamefont{and}
  \bibinfo{author}{\bibfnamefont{R.}~\bibnamefont{Pozzoli}},
  \bibinfo{journal}{Plasma Phys. Control. Fusion}
  \textbf{\bibinfo{volume}{30}}, \bibinfo{pages}{879} (\bibinfo{year}{1988}).

\bibitem[{\citenamefont{Farina and Pozzoli}(1989)}]{Farina1989}
\bibinfo{author}{\bibfnamefont{D.}~\bibnamefont{Farina}} \bibnamefont{and}
  \bibinfo{author}{\bibfnamefont{R.}~\bibnamefont{Pozzoli}},
  \bibinfo{journal}{Phys. Fluids B} \textbf{\bibinfo{volume}{1}},
  \bibinfo{pages}{815} (\bibinfo{year}{1989}).

\bibitem[{\citenamefont{Giruzzi et~al.}(1989)\citenamefont{Giruzzi, Fidone, and
  Meyer}}]{Giruzzi1989}
\bibinfo{author}{\bibfnamefont{G.}~\bibnamefont{Giruzzi}},
  \bibinfo{author}{\bibfnamefont{I.}~\bibnamefont{Fidone}}, \bibnamefont{and}
  \bibinfo{author}{\bibfnamefont{R.~L.} \bibnamefont{Meyer}},
  \bibinfo{journal}{Nucl. Fusion} \textbf{\bibinfo{volume}{29}},
  \bibinfo{pages}{1381} (\bibinfo{year}{1989}).

\bibitem[{\citenamefont{Dumont et~al.}(2000)\citenamefont{Dumont, Giruzzi, and
  Barbato}}]{Dumont2000}
\bibinfo{author}{\bibfnamefont{R.}~\bibnamefont{Dumont}},
  \bibinfo{author}{\bibfnamefont{G.}~\bibnamefont{Giruzzi}}, \bibnamefont{and}
  \bibinfo{author}{\bibfnamefont{E.}~\bibnamefont{Barbato}},
  \bibinfo{journal}{Phys. Plasmas} \textbf{\bibinfo{volume}{7}},
  \bibinfo{pages}{4972} (\bibinfo{year}{2000}).

\bibitem[{\citenamefont{Ando et~al.}(1986)\citenamefont{Ando, Ogura, Tanaka,
  Iida, Ide, Nakamura, Maekawa, Terumichi, and Tanaka}}]{Ando1986}
\bibinfo{author}{\bibfnamefont{A.}~\bibnamefont{Ando}},
  \bibinfo{author}{\bibfnamefont{K.}~\bibnamefont{Ogura}},
  \bibinfo{author}{\bibfnamefont{H.}~\bibnamefont{Tanaka}},
  \bibinfo{author}{\bibfnamefont{M.}~\bibnamefont{Iida}},
  \bibinfo{author}{\bibfnamefont{S.}~\bibnamefont{Ide}},
  \bibinfo{author}{\bibfnamefont{M.}~\bibnamefont{Nakamura}},
  \bibinfo{author}{\bibfnamefont{T.}~\bibnamefont{Maekawa}},
  \bibinfo{author}{\bibfnamefont{Y.}~\bibnamefont{Terumichi}},
  \bibnamefont{and} \bibinfo{author}{\bibfnamefont{S.}~\bibnamefont{Tanaka}},
  \bibinfo{journal}{Nucl. Fusion} \textbf{\bibinfo{volume}{26}},
  \bibinfo{pages}{107} (\bibinfo{year}{1986}).

\bibitem[{\citenamefont{Yamamoto et~al.}(1987)\citenamefont{Yamamoto, Hoshino,
  Kawashima, Uesugi, Mori, Suzuki, Ohta, Matoba, Kasai, Kawakami
  et~al.}}]{Yamamoto1987}
\bibinfo{author}{\bibfnamefont{Y.}~\bibnamefont{Yamamoto}},
  \bibinfo{author}{\bibfnamefont{K.}~\bibnamefont{Hoshino}},
  \bibinfo{author}{\bibfnamefont{H.}~\bibnamefont{Kawashima}},
  \bibinfo{author}{\bibfnamefont{Y.}~\bibnamefont{Uesugi}},
  \bibinfo{author}{\bibfnamefont{M.}~\bibnamefont{Mori}},
  \bibinfo{author}{\bibfnamefont{N.}~\bibnamefont{Suzuki}},
  \bibinfo{author}{\bibfnamefont{K.}~\bibnamefont{Ohta}},
  \bibinfo{author}{\bibfnamefont{T.}~\bibnamefont{Matoba}},
  \bibinfo{author}{\bibfnamefont{S.}~\bibnamefont{Kasai}},
  \bibinfo{author}{\bibfnamefont{T.}~\bibnamefont{Kawakami}},
  \bibnamefont{et~al.}, \bibinfo{journal}{Phys. Rev. Lett.}
  \textbf{\bibinfo{volume}{58}}, \bibinfo{pages}{2220} (\bibinfo{year}{1987}).

\bibitem[{\citenamefont{Maekawa et~al.}(1993)\citenamefont{Maekawa, Maehara,
  Minami, Kishigami, Kishino, Makino, Hanada, Nakamura, Terumichi, and
  Tanaka}}]{Maekawa1993}
\bibinfo{author}{\bibfnamefont{T.}~\bibnamefont{Maekawa}},
  \bibinfo{author}{\bibfnamefont{T.}~\bibnamefont{Maehara}},
  \bibinfo{author}{\bibfnamefont{T.}~\bibnamefont{Minami}},
  \bibinfo{author}{\bibfnamefont{Y.}~\bibnamefont{Kishigami}},
  \bibinfo{author}{\bibfnamefont{T.}~\bibnamefont{Kishino}},
  \bibinfo{author}{\bibfnamefont{K.}~\bibnamefont{Makino}},
  \bibinfo{author}{\bibfnamefont{K.}~\bibnamefont{Hanada}},
  \bibinfo{author}{\bibfnamefont{M.}~\bibnamefont{Nakamura}},
  \bibinfo{author}{\bibfnamefont{Y.}~\bibnamefont{Terumichi}},
  \bibnamefont{and} \bibinfo{author}{\bibfnamefont{S.}~\bibnamefont{Tanaka}},
  \bibinfo{journal}{Phys. Rev. Lett.} \textbf{\bibinfo{volume}{70}},
  \bibinfo{pages}{2561} (\bibinfo{year}{1993}).

\bibitem[{\citenamefont{Colborn et~al.}(1998)\citenamefont{Colborn, Squire,
  Porkolab, and Villase{\~n}or}}]{Colborn1998}
\bibinfo{author}{\bibfnamefont{J.~A.} \bibnamefont{Colborn}},
  \bibinfo{author}{\bibfnamefont{J.~P.} \bibnamefont{Squire}},
  \bibinfo{author}{\bibfnamefont{M.}~\bibnamefont{Porkolab}}, \bibnamefont{and}
  \bibinfo{author}{\bibfnamefont{J.}~\bibnamefont{Villase{\~n}or}},
  \bibinfo{journal}{Nucl. Fusion} \textbf{\bibinfo{volume}{38}},
  \bibinfo{pages}{783} (\bibinfo{year}{1998}).

\bibitem[{\citenamefont{Peysson and Imbeaux}(1999)}]{Peysson1999}
\bibinfo{author}{\bibfnamefont{Y.}~\bibnamefont{Peysson}} \bibnamefont{and}
  \bibinfo{author}{\bibfnamefont{F.}~\bibnamefont{Imbeaux}},
  \bibinfo{journal}{Rev. Sci. Instrum.} \textbf{\bibinfo{volume}{70}},
  \bibinfo{pages}{3987} (\bibinfo{year}{1999}).

\bibitem[{\citenamefont{Giruzzi et~al.}(2000)\citenamefont{Giruzzi, Darbos,
  Dumont, Magne, Peysson, Zou, Bouquey, Courtois, Hoang, Imbeaux
  et~al.}}]{Giruzzi2000}
\bibinfo{author}{\bibfnamefont{G.}~\bibnamefont{Giruzzi}},
  \bibinfo{author}{\bibfnamefont{C.}~\bibnamefont{Darbos}},
  \bibinfo{author}{\bibfnamefont{R.}~\bibnamefont{Dumont}},
  \bibinfo{author}{\bibfnamefont{R.}~\bibnamefont{Magne}},
  \bibinfo{author}{\bibfnamefont{Y.}~\bibnamefont{Peysson}},
  \bibinfo{author}{\bibfnamefont{X.}~\bibnamefont{Zou}},
  \bibinfo{author}{\bibfnamefont{F.}~\bibnamefont{Bouquey}},
  \bibinfo{author}{\bibfnamefont{L.}~\bibnamefont{Courtois}},
  \bibinfo{author}{\bibfnamefont{G.}~\bibnamefont{Hoang}},
  \bibinfo{author}{\bibfnamefont{F.}~\bibnamefont{Imbeaux}},
  \bibnamefont{et~al.}, in \emph{\bibinfo{booktitle}{Proceedings of the
  Eighteenth Conference on Plasma Physics and Controlled Nuclear Fusion
  Research}} (\bibinfo{publisher}{International Atomic Energy Agency},
  \bibinfo{address}{Vienna}, \bibinfo{year}{2000}), \bibinfo{note}{paper
  IAEA-CN-77/EXP4/02}.

\bibitem[{\citenamefont{Pericoli-Ridolfini
  et~al.}(2001)\citenamefont{Pericoli-Ridolfini, Barbato, Bruschi, Dumont,
  Gandini, Giruzzi, Gormezano, Granucci, Panaccione, Peysson
  et~al.}}]{Pericoli2001}
\bibinfo{author}{\bibfnamefont{V.}~\bibnamefont{Pericoli-Ridolfini}},
  \bibinfo{author}{\bibfnamefont{E.}~\bibnamefont{Barbato}},
  \bibinfo{author}{\bibfnamefont{A.}~\bibnamefont{Bruschi}},
  \bibinfo{author}{\bibfnamefont{R.}~\bibnamefont{Dumont}},
  \bibinfo{author}{\bibfnamefont{F.}~\bibnamefont{Gandini}},
  \bibinfo{author}{\bibfnamefont{G.}~\bibnamefont{Giruzzi}},
  \bibinfo{author}{\bibfnamefont{C.}~\bibnamefont{Gormezano}},
  \bibinfo{author}{\bibfnamefont{G.}~\bibnamefont{Granucci}},
  \bibinfo{author}{\bibfnamefont{L.}~\bibnamefont{Panaccione}},
  \bibinfo{author}{\bibfnamefont{Y.}~\bibnamefont{Peysson}},
  \bibnamefont{et~al.}, in \emph{\bibinfo{booktitle}{RF Power in Plasmas}},
  edited by \bibinfo{editor}{\bibfnamefont{T.~K.} \bibnamefont{Mau}}
  \bibnamefont{and} \bibinfo{editor}{\bibfnamefont{J.}~\bibnamefont{deGrassie}}
  (\bibinfo{publisher}{AIP}, \bibinfo{address}{Meville, NY},
  \bibinfo{year}{2001}), p. \bibinfo{pages}{225}.

\bibitem[{\citenamefont{Antonsen and Chu}(1982)}]{Antonsen1982}
\bibinfo{author}{\bibfnamefont{T.~M.} \bibnamefont{Antonsen}} \bibnamefont{and}
  \bibinfo{author}{\bibfnamefont{K.~R.} \bibnamefont{Chu}},
  \bibinfo{journal}{Phys. Fluids} \textbf{\bibinfo{volume}{25}},
  \bibinfo{pages}{1295} (\bibinfo{year}{1982}).

\bibitem[{\citenamefont{Risken}(1989)}]{Risken1989}
\bibinfo{author}{\bibfnamefont{H.}~\bibnamefont{Risken}},
  \emph{\bibinfo{title}{The Fokker-Planck equation}}
  (\bibinfo{publisher}{Springer-Verlag}, \bibinfo{address}{Berlin},
  \bibinfo{year}{1989}), \bibinfo{edition}{2nd} ed.

\bibitem[{\citenamefont{Brambilla}(1998)}]{Brambilla1998}
\bibinfo{author}{\bibfnamefont{M.}~\bibnamefont{Brambilla}},
  \emph{\bibinfo{title}{Kinetic Theory of Plasma Waves}}
  (\bibinfo{publisher}{Clarendon Press}, \bibinfo{address}{Oxford},
  \bibinfo{year}{1998}).

\bibitem[{\citenamefont{Rax and Moreau}(1989)}]{Rax1989}
\bibinfo{author}{\bibfnamefont{J.~M.} \bibnamefont{Rax}} \bibnamefont{and}
  \bibinfo{author}{\bibfnamefont{D.}~\bibnamefont{Moreau}},
  \bibinfo{journal}{Nucl. Fusion} \textbf{\bibinfo{volume}{29}},
  \bibinfo{pages}{1751} (\bibinfo{year}{1989}).

\bibitem[{\citenamefont{Cadjan and Ivanov}(1999)}]{Cadjan1999}
\bibinfo{author}{\bibfnamefont{M.~G.} \bibnamefont{Cadjan}} \bibnamefont{and}
  \bibinfo{author}{\bibfnamefont{M.~F.} \bibnamefont{Ivanov}},
  \bibinfo{journal}{J. Plasma Phys.} \textbf{\bibinfo{volume}{61}},
  \bibinfo{pages}{89} (\bibinfo{year}{1999}).

\bibitem[{\citenamefont{Castej\'on and Eguilior}(2003)}]{Castejon2003}
\bibinfo{author}{\bibfnamefont{F.}~\bibnamefont{Castej\'on}} \bibnamefont{and}
  \bibinfo{author}{\bibfnamefont{S.}~\bibnamefont{Eguilior}},
  \bibinfo{journal}{Plasma Phys. Control. Fusion}
  \textbf{\bibinfo{volume}{45}}, \bibinfo{pages}{159} (\bibinfo{year}{2003}).

\bibitem[{\citenamefont{Garc\'{\i}a-Palacios and
  L\'azaro}(1998)}]{GarciaPalacios1998}
\bibinfo{author}{\bibfnamefont{J.~L.} \bibnamefont{Garc\'{\i}a-Palacios}}
  \bibnamefont{and} \bibinfo{author}{\bibfnamefont{F.~J.}
  \bibnamefont{L\'azaro}}, \bibinfo{journal}{Phys. Rev. B}
  \textbf{\bibinfo{volume}{58}}, \bibinfo{pages}{14937} (\bibinfo{year}{1998}).

\bibitem[{\citenamefont{Greiner et~al.}(1988)\citenamefont{Greiner,
  Strittmatter, and Honerkamp}}]{Greiner1988}
\bibinfo{author}{\bibfnamefont{A.}~\bibnamefont{Greiner}},
  \bibinfo{author}{\bibfnamefont{W.}~\bibnamefont{Strittmatter}},
  \bibnamefont{and}
  \bibinfo{author}{\bibfnamefont{J.}~\bibnamefont{Honerkamp}},
  \bibinfo{journal}{J. Stat. Phys.} \textbf{\bibinfo{volume}{51}},
  \bibinfo{pages}{95} (\bibinfo{year}{1988}).

\bibitem[{\citenamefont{Bonoli and Englade}(1986)}]{Bonoli1986}
\bibinfo{author}{\bibfnamefont{P.~T.} \bibnamefont{Bonoli}} \bibnamefont{and}
  \bibinfo{author}{\bibfnamefont{R.~C.} \bibnamefont{Englade}},
  \bibinfo{journal}{Phys. Fluids} \textbf{\bibinfo{volume}{29}},
  \bibinfo{pages}{2937} (\bibinfo{year}{1986}).

\end{thebibliography}


\newpage

\newpage

\newpage

\newpage

\newpage
\begin{figure}[htbp]
\centering
\includegraphics[width=11cm]{Figs/langevin-indv-iv4}
\caption{Same as Fig.~\ref{fig:langevin-indv-iv1} but for initial position $(u_{\parallel 0},u_{\perp 0})=(6,0)$.}\label{fig:langevin-indv-iv4} 
\end{figure}

\newpage

\newpage

\newpage

\newpage

\newpage

\newpage

\newpage

\newpage

\newpage

\end{document}